\documentclass[twocolumn,english,prb,showpacs]{revtex4-1}

\usepackage{float}
\usepackage[dvips,final]{graphicx}
\usepackage{color}
\usepackage{amsmath}
\usepackage{graphicx, subfigure}
\usepackage{pslatex}
\usepackage{relsize}
\usepackage{bm}
\usepackage{verbatim}

\usepackage{appendix}

\usepackage{booktabs}
\usepackage{xcolor}
\usepackage{soul}

\usepackage{natbib}

\begin{document}
\title{Computational screening of Fe-Ta hard magnetic phases}

\author{S. Arapan$^{1}$}
\email{Corresponding author: sergiu.arapan@gmail.com}

\author{P. Nieves$^{1,2}$}

\author{H. C. Herper$^{3}$}
\author{D. Legut$^{1}$}

\affiliation{$^1$ IT4Innovations, V\v{S}B - Technical University of Ostrava, 17. listopadu 15, 70833 Ostrava-Poruba, Czech Republic}

\affiliation{$^2$ ICCRAM, International Research Center in Critical Raw Materials and Advanced Industrial Technologies, Universidad de Burgos, 09001 Burgos, Spain}

\affiliation{$^3$ Department of Physics and Astronomy, Uppsala University, Box 516, 75121 Uppsala, Sweden}

\date{\today}

\begin{abstract}
In this work we perform a systematic calculation of the Fe-Ta phase diagram to discover novel hard magnetic phases. By using structure prediction methods based on evolutionary algorithms, we identify two new energetically stable magnetic structures: a tetragonal Fe$_3$Ta (space group 122) and cubic Fe$_5$Ta (space group 216) binary phases. The tetragonal structure is estimated to have both high saturation magnetization ($\mu_0$M$_s$=1.14 T) and magnetocrystalline anisotropy (K$_1$=2.17 MJ/m$^3$) suitable for permanent magnet applications. The high-throughput screening of magneto-crystalline anisotropy also reveals two low energy metastable hard magnetic phases: Fe$_5$Ta$_2$ (space group 156) and Fe$_{6}$Ta (space group 194), that may exhibit intrinsic magnetic properties comparable to SmCo$_5$ and Nd$_2$Fe$_{14}$B, respectively.
\end{abstract}
\pagebreak
\pacs{31.15.A-,75.50.Ww, 75.30.Gw, 07.05.Tp }
\maketitle

\section{Introduction}

Many technological applications used for information storage and green energy generation, like motors for hybrid and electric cars and direct-drive wind turbines, rely on high quality permanent magnets (PMs) \cite{Scheu}.  The increasing importance of PMs in modern society has resulted in a renewed interest in the design of new magnet materials that are cheaper and contain less critical components like Rare-Earth (RE). One possible alternative to RE-PM could be RE-free Fe(Co)-rich intermetallic compounds. To be viewed as a good PM a ferromagnetic compound must have a high Curie temperature ($T_C>$ 400 K), a high saturation magnetization ($\mu_0M_S>$ 1T) and a uniaxial anisotropy energy larger than 1 MJ/m$^3$, since large anisotropy is a key factor for the large coercivity needed for high-performance PMs\cite{SKOKOV2018289}. The main contribution to the magnetic anisotropy is usually magnetocrystalline (K), that is, a combined effect of crystal-field splitting (or band formation) and spin-orbit coupling. This mechanism is also responsible for the surface, interface and magnetostrictive anisotropies. Since none of the 4d and 5d dopants is ferromagnetic at room temperature, the 3d sublattice must spin-polarize the partially filled 4d/5d shells. This ensures a net spin-orbit effect, as required for the creation of the anisotropy. Therefore, the addition of heavy elements for a large spin-orbit coupling, as Hf, Ta, Bi, Sn or Zr, to the Fe-Co alloys could form hard magnetic phases suitable for PM applications with a low raw materials cost. However, a large magnetocrystalline anisotropy in Fe-based alloys can only be found in non-cubic uniaxial structures, like FePt where a large K=7 MJ/m$^3$ is observed in the tetragonal L1$_0$ structure.\cite{Ivanov1973} Therefore, the theoretical research of such compounds should be focused on non-cubic uniaxial structures as tetragonal, hexagonal or rhombohedral.

In Fe-Co alloys, Ta can induce some interesting features. For instance, recent gradient-composition sputtering experiments made by Phuoc \textit{et al.}\cite{Phuoc1, Phuoc2} showed that Co-Fe-Ta exhibits a peculiar increased magnetic anisotropy with temperature much larger than Fe-Co-M where M~$=$~Hf, Zr, Lu. An improved coercivity has been reported  in magnets such as (Fe,Co)$_2$B and Ce(Co,Fe,Cu)$_5$ after the incorporation of small amount of Ta  \cite{Taskaev,Lami2019}. While the Co-Ta phase diagram shows up to seven stable binary phases\cite{Villars2014,SHINAGAWA201487}, only two structures have been found in the phase diagram of Fe-Ta: Laves phase (C14) Fe$_2$Ta space group (SG) 194, P6$_3$/mmc and $\mu$-phase FeTa SG 166, R-3m. The Fe$_2$Ta structure is a paramagnet in which either Fe or Ta excess can induce a ferromagnetic ordering at low temperatures ($\lesssim 150$ K) \cite{Kai}. Theoretical calculations show an easy cone magnetocrystalline anisotropy in the ferromagnetic state\cite{Edstrom}. Recently, Gabay \textit{et al.} reported melt-spun alloys made of Fe-rich Fe$_2$Ta and Fe-bcc without sufficiently high coercivities (around to 0.5 kOe at room-temperature) for PM applications\cite{Fe2Ta_George}. The other known stable crystalline phase, $\mu$-phase FeTa, is an antiferromagnet with N\'{e}el temperature around 336 K\cite{Ahmed}. Additionally, small amount of Ta can also be inserted into the Fe-bcc structure\cite{Villars2014}. In the case of structurally amorphous systems, Fe$_{9}$Ta thin films have been found to exhibit characteristics of a soft ferromagnetic material with very low coercivities (1$-$10 mT), saturation magnetization around 2 T and Extraordinary Hall Effect \cite{SHAJI}.

In this work we performed a systematic computational exploration of Fe-rich Fe-Ta compounds to find new magnetic structures with intrinsic properties suitable for high-performance PMs. Theoretical search for new magnetic phases is done in few stages. First, we searched for stable and low-energy metastable structures for different Fe-rich Fe$_{1-x}$Ta$_x$ binaries. Details of calculations and results of the crystal phase exploration are presented in the Section~\ref{section:crystal_str_pred}. Second, a set of structures with intrinsic magnetic properties fulfilling the criteria of a performance PM, i. e., exhibiting negative enthalpy of formation, $\Delta H <0$, high saturation magnetization and uniaxial lattice, are selected from the collection of predicted structures as well as from the AFLOW database~\cite{Aflow_1}. We calculate the magnetocrystalline anisotropy (MCA) of all of these phases and identify a smaller subset of structures, which exhibit intrinsic properties of hard magnets. These few phases are analyzed in more detail to understand the possible mechanisms of a high MCA in RE-free intermetallic compounds. We present the corresponding results and calculation details in Section~\ref{section:MAE} and Appendix \ref{App_A}. At the third stage, we include the effects of finite temperature on the phase stability and performed calculations of the exchange integrals of the few selected structures to estimate the Curie temperature (T$_C$), thus, having screened all considered Fe$_{1-x}$Ta$_x$ binaries according to all three criteria to select a structure as a promising PM. Calculation detail and results of this stage are presented in Section~\ref{section:Temp_effec}. We finalized our work by performing a study of a possible stabilization of metastable phases as thin films by epitaxial growth on suitable substrates. We performed calculations of the elastic properties of selected structures and results are given in Section~\ref{section:sub}. The paper is completed by a Conclusions section.        

\section{Crystal phase space exploration}
\label{section:crystal_str_pred}

\begin{figure}[ht!]
\centering
\includegraphics[width=\columnwidth ,angle=0]{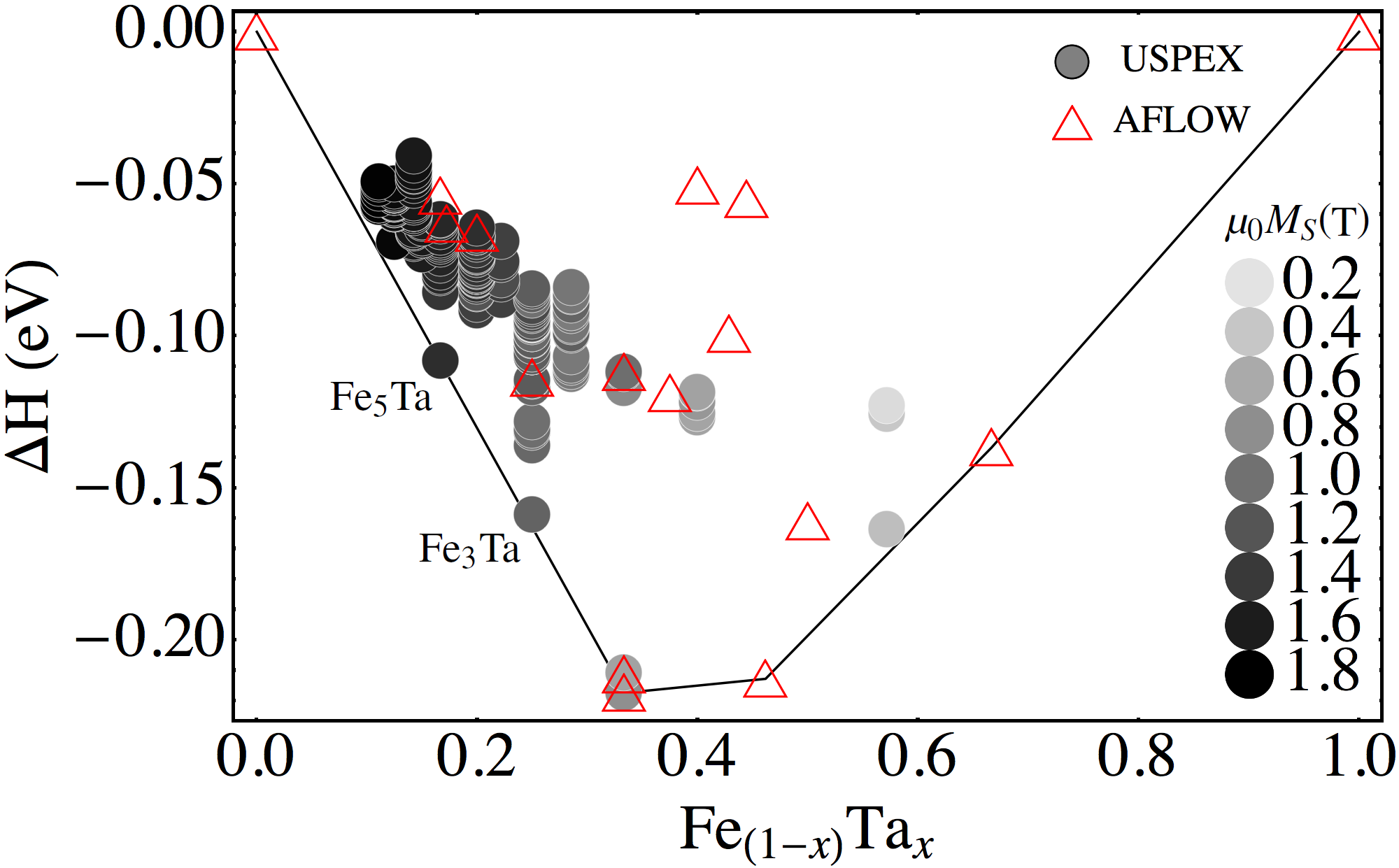}
\caption{Convex hull diagram of Fe-Ta showing phases predicted by using the USPEX (disks) and the reference structures from the AFLOW database (hollow red triangles). The magnitude of saturation magnetization is represented by the intensity of the gray scale.}
\label{fig:phase_ms}
\end{figure}

We began our study by exploring the phase space of ordered Fe-rich Fe$_{1-x}$Ta$_x$ binaries for several compositions (Fe$_2$Ta, Fe$_3$Ta, Fe$_4$Ta, Fe$_5$Ta, Fe$_5$Ta$_2$, Fe$_6$Ta, Fe$_7$Ta, Fe$_7$Ta$_2$, Fe$_{17}$Ta$_3$, Fe$_8$Ta) using crystal predicting methods. We searched for possible structures by using the USPEX software~\cite{uspex,uspex_web}, an implementation of the evolutionary algorithm, and the Vienna Ab Initio Simulation Package (VASP) \cite{vasp_1,vasp_2,vasp_3}. As in our recent work~\cite{Arapan_AGA_1}, we used an optimized USPEX-VASP interface for the efficient computational search of magnetic structures. We have run the USPEX with the evolutionary algorithm method for 3-D structures and choosing the enthalpy as a fitness criterion. The population size was set to be twice the number of atoms in the system and the maximum number of generations to be calculated was set to 40. We used 15 generations for convergence and best 65\% of the population size was used for new generation. Out of the new structures 45\% of  were obtained by heredity, 5\% by soft-mutations, 5\% by lattice mutations, and 45\% were randomly generated. For the random generation, we used all space groups except triclinic and monoclinic lattice systems. All of the VASP calculations were done with the projector augmented wave (PAW) method \cite{PAW} and the generalized gradient approximation (GGA) of Perdew, Burke, and Ernzerhof (PBE)~\cite{PBE} to the exchange correlation part of the energy functional (PAW PBE potentials version 5.4). We performed calculations with the $p$ semi-core electrons treated as valence ones. Best generated structures were fully relaxed until maximum force component became less than 5$\times$10$^{-3}$\,eV/\AA. We used an automatic k-points generating scheme with the length $l=40$ and an energy cut-off up to 1.4 of the default VASP energy cut-off. Additionally, we also ran calculations for a set of low-energy Fe-Ta phases available in the AFLOW database~\cite{Aflow_1,Aflow_2}, which we use for the reference. We present relevant results of this study in Fig.~\ref{fig:phase_ms}. This figure shows the convex hull diagram of the Fe-Ta binary system. The hull, shown by solid black lines is formed by using calculated energies of Fe, Ta, FeTa and Fe$_2$Ta phases. Symbols correspond to the values of enthalpy of formation $\Delta H$ of various phases with respect to single elements Fe and Ta as: 
\begin{equation}
\Delta H(Fe_nTa_m)=E(Fe_nTa_m)-n\cdot E(Fe)-m\cdot E(Ta),
\end{equation}
where $E(.)$ is the energy at equilibrium conditions ($PV=0$), $n$ and $m$ are the number of atoms of Fe and Ta in the formula unit of the Fe$_n$Ta$_m$ compound, respectively. $E(Fe)$ and $E(Ta)$ correspond to the energy of computationally optimized bcc phases of Fe and Ta. The relative position of these symbols with respect to the hull provides the information of the energy stability of structures at $T=0K$. Our search reveals a set of new metastable magnetic phases (shown by filled disks) with two structures, a tetragonal Fe$_3$Ta (SG 122) and a cubic Fe$_5$Ta (SG 216), in the proximity of the convex hull. In addition, we show in Fig.~\ref{fig:phase_ms} the magnitude of saturation magnetization of calculated phases, represented by the intensity of the gray scale. We can observe that all ferromagnetic Fe-Ta compounds with Fe content above 75 at.\% have $\mu_0M_S\gtrsim 1$. In this respect, the new tetragonal Fe$_3$Ta also exhibits a saturation magnetization, which qualifies it as a promising PM structure. While the new energetically stable Fe$_5$Ta phase is not suitable as a PM structure because of its cubic symmetry, it is interesting to analyze this phase in view of possible experimental synthesis as a validation of our theoretical predictions. More details of the Fe-Ta phases calculated with USPEX can be found in the Novamag database\cite{novamag,Nieves_2019}.   

\begin{table*}[ht]
\centering
\caption{Space group, enthalpy of formation, saturation magnetization, MCA constants, lattice parameters, Curie temperature and magnetic hardness parameter of main Fe-Ta phases discussed in this work. Data of SmCo$_5$ and Nd$_2$Fe$_{14}$B are also shown for comparison. The chemical formula of known stable structures are written in bold and superscripts stand for references.}
\label{data_table}
\begin{tabular}{cccccccccccc}
Compound    &  \begin{tabular}[c]{@{}c@{}}Space \\ group\end{tabular} & \begin{tabular}[c]{@{}c@{}}$\Delta$H\\ (eV/atom)\end{tabular} & \begin{tabular}[c]{@{}c@{}}$\mu_0$M$_s$\\ (T)\end{tabular} & \begin{tabular}[c]{@{}c@{}} K$_1$\\ (MJ/m$^3$)\end{tabular} & \begin{tabular}[c]{@{}c@{}} K$_2$\\ (MJ/m$^3$)\end{tabular} & \begin{tabular}[c]{@{}c@{}}a\\ \:\:\:($\AA$)\:\:\:\end{tabular} & \begin{tabular}[c]{@{}c@{}}b\\ \:\:\:($\AA$)\:\:\:\end{tabular} & \begin{tabular}[c]{@{}c@{}}c\\ \:\:\:($\AA$)\:\:\:\end{tabular} & \begin{tabular}[c]{@{}c@{}}Temp\\(K)\end{tabular}&
\begin{tabular}[c]{@{}c@{}}T$_C$\\(K)\end{tabular} & $\kappa$ \\ \hline
\textbf{Fe$_2$Ta}    & 194 & -0.2053 & 0.63 & -0.73 & 1.48 &  4.776                                                          & 4.776                                                           & 7.824  & 0 & -   & -                                                 
\\
\textbf{Fe$_2$Ta}\cite{Edstrom}     & 194 & -0.2350 & 0.66 & -0.27 &  1.52                                                      & 4.811                                                           & 4.811 & 7.874   & 0 & - & -                                                                                                     
\\ \hline

Fe$_3$Ta     & 122 & -0.1588 & 1.14 & 2.17 &  -0.84 & 6.733                                                           & 6.733                                                           & 13.455  & 0 & 364 & 1.45                                                      
\\
Fe$_5$Ta     & 216 & -0.1083 & 1.53 & - &  -                                                      & 6.678                                                           & 6.678 & 6.678   & 0 & - & -                                                                                                         
\\ \hline
Fe$_5$Ta$_2$     & 156 & -0.1143 & 1.00 & 10.43 &  6.22 & 4.713                                                           & 4.713                                                          &  4.744 & 0 & 724   & 3.62                                                
 \\
Fe$_6$Ta  & 194 & -0.0370                                                           & 1.62 & 5.77 &  0.60 & 4.627                                                           & 4.627                                                           & 9.353 & 0 & 886   & 1.66                                                    
\\ \hline
\textbf{SmCo$_5$}\cite{Coeybook}    & 191 & - & 1.08 & 17.20 &  - & 4.990 & 4.990 & 3.980 & 300 & 1020 & 4.30                                                   
 \\ 
\textbf{Nd$_2$Fe$_{14}$B}\cite{Coeybook}    & 136 & - & 1.61 & 4.90 &  - & 8.790 & 8.790 & 12.180 & 300 & 588 & 1.54
\\ \hline
\end{tabular}
\end{table*}


\begin{table*}[ht]
\centering
\caption{Crystallographic data, spin magnetic moment ($\mu_{spin}$), orbital magnetic moment ($\mu_{orb}$) and SOC energy (E$_{so}$) of Fe$_5$Ta$_2$ (SG 156).}
\label{data_table2a}
\begin{tabular}{@{}ccccccccccccc@{}}
Compound     & Atom & \begin{tabular}[c]{@{}c@{}}Wyckoff \\ position\end{tabular} & x     & y     & z      & \begin{tabular}[c]{@{}c@{}}$\mu_{spin}${[}001{]}\\ $(\mu_B)$\end{tabular} &
\begin{tabular}[c]{@{}c@{}}$\mu_{orb}${[}001{]}\\ $(\mu_B)$\end{tabular} &
\begin{tabular}[c]{@{}c@{}}$\mu_{spin}${[}100{]}\\ $(\mu_B)$\end{tabular} &
\begin{tabular}[c]{@{}c@{}}$\mu_{orb}${[}100{]}\\ $(\mu_B)$\end{tabular} &
\begin{tabular}[c]{@{}c@{}}E$_{so}${[}100{]}\\ (meV)\end{tabular} & \begin{tabular}[c]{@{}c@{}}E$_{so}${[}001{]}\\ (meV)\end{tabular} & \begin{tabular}[c]{@{}c@{}}$\Delta$E$_{so}$\\ (meV)\end{tabular} \\ \midrule
Fe$_5$Ta$_2$ & Fe$_1$   & 1a                                                          & 0   & 0   & 0.514  & 1.456  & 0.189 &  1.442 & 0.077                                                        & -17.403                                                           & -20.441                                                           & 3.038                                                             \\
             & Fe$_2$   & 1b                                                          & 1/3   & 2/3   & 0.746  & 2.353 & 0.206     & 2.358 &  0.095                                                      & -17.395                                                           & -19.373                                                           & 1.978                                                             \\
             & Fe$_3$   & 3d                                                          & 0.498 & 0.502 & 0.258  & 1.707 & 0.111      &   1.702  &    0.067                                                 & -16.559                                                           & -17.926                                                           & 1.367                                                             \\
             & Ta$_1$   & 1c                                                          & 2/3   & 1/3   & 0.741  & -0.536  & 0.072  &  -0.537 &   0.028                                                     & -288.187                                                          & -289.123                                                          & 0.936                                                             \\
             & Ta$_2$   & 1a                                                          & 0   & 0   & 0    & -0.524   & 0.052 &  -0.516 &    0.027                                                    & -297.208                                                          & -300.756                                                          & 3.548                                                         \\ \bottomrule
\end{tabular}
\end{table*}


\section{Screening of magneto-crystalline anisotropy}
\label{section:MAE}  

At the next stage of our study, we have performed calculations of magnetocrystalline anisotropy energy (MAE) on a set of selected Fe$_{1-x}$Ta$_x$ binary structures. We screened a set of structures consisting of phases predicted by the USPEX search and phases available in the AFLOW database to select a subset of structures with $\Delta H<0$, $\mu_0M_S\gtrsim 1$ and uniaxial lattice system (tetragonal, rhombohedral and hexagonal). We calculated the MAE by performing VASP non-colinear spin-polarized calculations in a high-throughput manner~ \cite{Nieves_2019}. Calculations of MAE require, usually, a higher accuracy, and, especially, a denser k-mesh to sample the reciprocal space. This, inevitably, increases the amounts of computational time and memory. To decrease the computational demand, we used the PAW PBE potential with minimum number of valence electrons. Extended test calculations showed that the addition of the $p$ semi-core electrons do not change considerably calculated values (some detailed examples of the MAE calculations are provided in Appendix \ref{App_A}). For all of the MAE calculations we used an energy cut-off 1.50 times larger than the default one (ENCUT~$=401.823$\,eV) and the energy of a system was calculated with a tolerance EDDIF=$10^{-9}$\,eV. We also found that an automatic k-point mesh with the length $l=60$ is enough to provide reliable MAE. The MAE was calculated as energy difference between the configurations with different colinear spin arrangements:   
\begin{equation}
\Delta E=E_{\theta=0}-E_{\theta},
\label{eq:E_mae}
\end{equation}  
where $\theta$ is the angle between the direction of spins and the $z-$ axis. The energy of a given $\theta$ configuration, $E_{\theta}$, was calculated in a non-self-consistent way by using the charge density and wavefunctions of a colinear spin-polarized calculation. We estimated the anisotropy constants $K_1$ and $K_2$ for uniaxial systems by fitting the MAE to the following equation:
\begin{equation}
\Delta E(\theta)=K_1 sin^2( \theta) + K_2 sin^4(\theta).
\label{eq:K1K2}
\end{equation}

\begin{figure}[ht!]
\centering
\includegraphics[width=0.9\columnwidth ,angle=0]{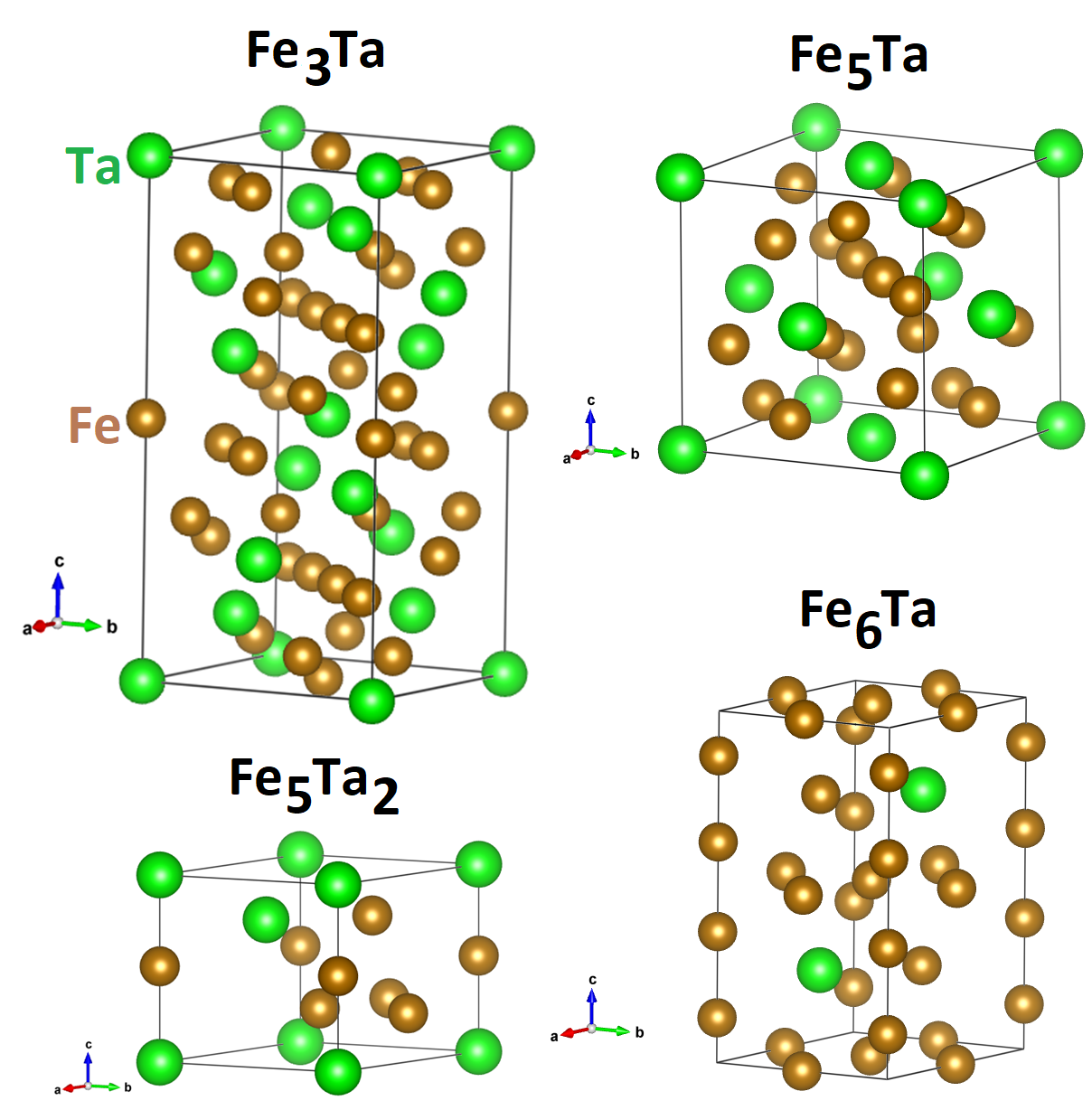}
\caption{Unit cells of four interesting Fe-Ta structures studied in this work: Fe$_3$Ta (SG 122), Fe$_5$Ta (SG 216), Fe$_6$Ta (SG 194) and Fe$_5$Ta$_2$ (SG 156).}
\label{fig:unit_cells}
\end{figure}

\begin{figure}[ht!]
\centering
\includegraphics[width=\columnwidth ,angle=0]{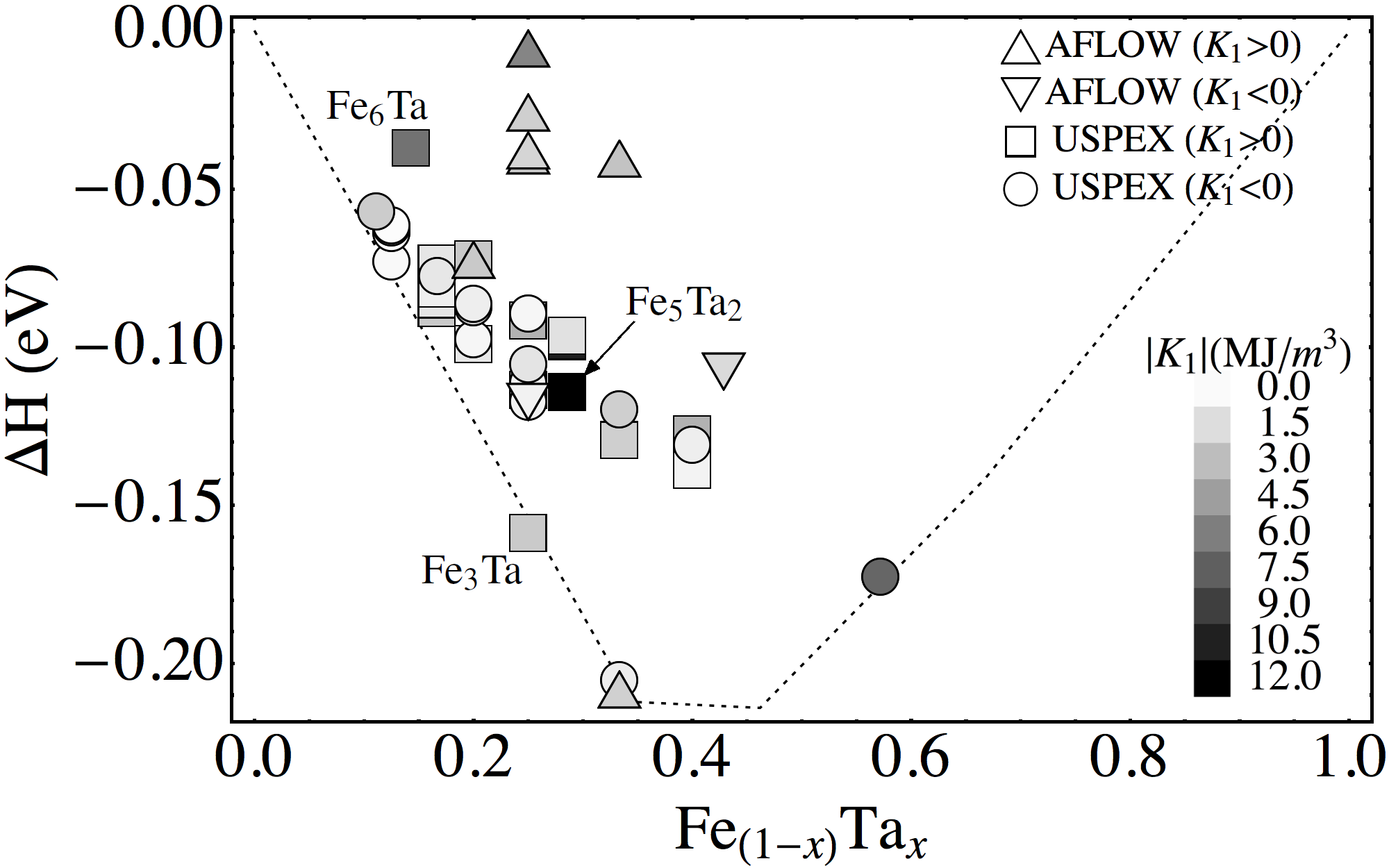}
\caption{Convex hull diagram of set of uniaxial Fe-Ta phases selected from the USPEX predicted phases and the AFLOW database. Gray scale represents the magnitude of the first MCA constant $K_1$. Various symbols correspond to the sign of $K_1$.}
\label{fig:phase_k}
\end{figure}

In Fig.~\ref{fig:phase_k} we show the convex hull diagram of uniaxial Fe-Ta phases obtained by performing the MAE calculations, where the magnitude of the first MCA constant $K_1$ is represented by the intensity of the gray scale and its sign by a given symbol. This figure reveals two hard magnetic phases: Fe$_5$Ta$_2$ (SG 156) with $\mu_0M_S=1$ T, $K_1+K_2=16.65$ MJ/m$^3$ and magnetic hardness parameter $\kappa=\sqrt{K_1/(\mu_0M_S^2)}=3.62$,  and Fe$_{6}$Ta (SG 194) with $\mu_0M_S=1.62$ T, $K_1=5.77$ MJ/m$^3$ and $\kappa=1.66$. We can see that these values are comparable to the state of the art RE-PM SmCo$_5$ ($\mu_0M_S=1.08$ T, $K_1=17.2$ MJ/m$^3$) and Nd$_2$Fe$_{14}$B ($\mu_0M_S=1.08$ T, $K_1=4.9$ MJ/m$^3$) at room temperature, respectively\cite{Coeybook}. Here, we also highlight Fe$_3$Ta (SG 122) which combines both high phase stability (lays on the enthalpy convex hull) and good magnetic properties ($\mu_0M_S=1.14$ T and $K_1=2.17$ MJ/m$^3$). In Fig.~\ref{fig:unit_cells} we show the unit cells of the above mentioned structures: two new energetically stable phases Fe$_3$Ta (SG 122) and Fe$_5$Ta (SG 216), and two low energy metastable hard magnetic phases Fe$_5$Ta$_2$ (SG 156) and Fe$_6$Ta (SG 194). In Tables \ref{data_table}, \ref{data_table2a} and  Appendix \ref{App_B} we also provide various structural and magnetic properties of these phases. In the following sections we analyze these four structures in more detail. To some extent, these results may also be applied to Fe-Nb or Fe-Ta-Nb systems due to the similarity between Ta and Nb. For instance, Fe$_5$Nb$_2$ (SG 156) shows a large easy axis MAE with $K_1=7.7$ MJ/m$^3$ and $K_2=2.2$ MJ/m$^3$.

\begin{figure}[ht!]
\centering
\includegraphics[width=\columnwidth,angle=0]{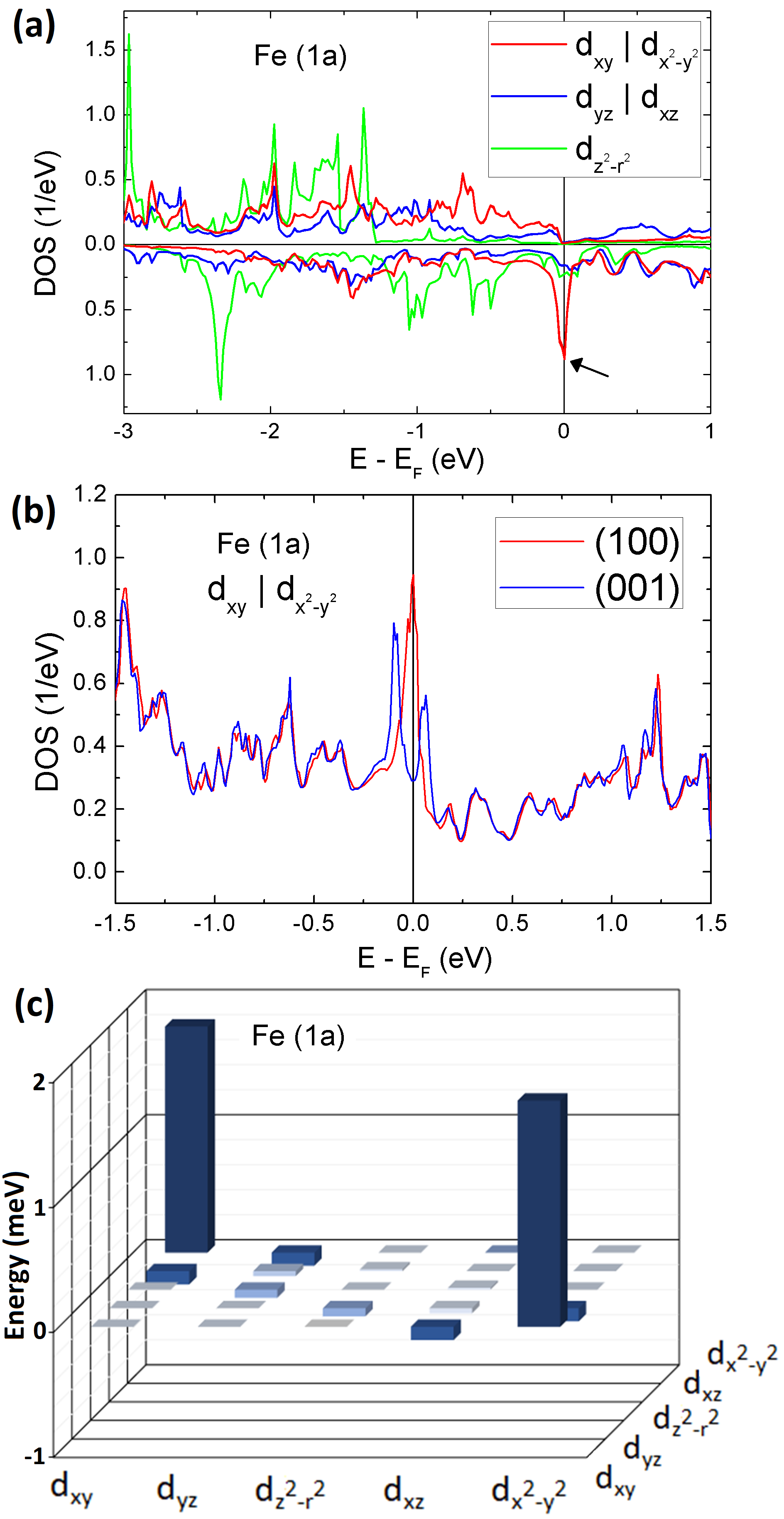}
\caption{a) Partial DOS projected on d states of Fe atom at Wyckoff position (1a) of Fe$_5$Ta$_2$ without SOC interaction. b) Total DOS projected on d$_{xy}$ and d$_{x^2-y^2}$ states of Fe (1a) in Fe$_5$Ta$_2$ when the magnetization is aligned along the easy axis (blue line) and hard plane (red line) including SOC interaction. c) Change of the SOC matrix elements of Fe (1a) in Fe$_5$Ta$_2$, when magnetization goes from hard plane [100] to easy axis [001].}
\label{fig:mae_Fe5Ta2_Fe1_Ta_1}
\end{figure}

In order to identify the source of the large MAE found in Fe$_5$Ta$_2$ we follow a similar analysis as in Refs.~\onlinecite{antropov2014,Liu2019}. Namely, we analyzed the spin-orbit coupling energy of each atom with all spin orientation along z and x axis, $E_{so}[001]$ and $E_{so}[100]$, see Table \ref{data_table2a}. We observe that main contribution to total MAE comes from Fe and Ta atoms at the Wyckoff (1a) site with $\Delta E_{so}=E_{so}[100]-E_{so}[001]=3.04$ meV and $3.55$ meV, respectively.  Next, we try to find the electronic states responsible for the this large MAE. To do so, we analyzed the partial density of states (DOS) projected on d states of Fe (1a) without  spin-orbit coupling (SOC) interaction, see Fig.~\ref{fig:mae_Fe5Ta2_Fe1_Ta_1}(a). It shows that there is a large DOS in the minority spin channel of d$_{xy}$ and d$_{x^2-y^2}$ states (d orbitals that lay on the hard plane) right at the Fermi level. This peak is decomposed into two smaller peaks below and above the Fermi level when the SOC is included and the magnetization is aligned along the easy axis [001], Fig.~\ref{fig:mae_Fe5Ta2_Fe1_Ta_1}(b), decreasing the total energy and inducing a large MAE. The fact that the coupling between the minority spin channel gives the largest contribution to the SOC is also supported by the maximum values of the orbital magnetic moments in the easy direction of magnetization (see Table \ref{data_table2a})\cite{Edstrom}. In this process, where the magnetization goes from hard plane to easy axis, the SOC matrix element $\langle d_{xy}|\hat{H}_{so}| d_{x^2-y^2}\rangle$ exhibits a much greater change than the other ones, Fig.~\ref{fig:mae_Fe5Ta2_Fe1_Ta_1}(c). It is interesting to note that Fe$_5$Ta$_2$ is capable to transfer very efficiently its spin-orbit coupling energy to total MAE with a ratio $\Delta E_{so}/\Delta E=1.41$, where $\Delta E=E[100]-E[001]=9.55$ meV is the total energy difference between the states with all spins orientated along [100] and [001] direction. For instance, this spin-orbit reduction factor is smaller than in the L1$_0$ FePt (1.84) and CoPt (1.67)\cite{antropov2014}. Typically, when the second order perturbation theory holds, this factor is close to 2.
This analysis may also explain the dependence of calculated MAE with energy smearing methods shown in Fig.~\ref{fig:mae_Fe5Ta2} (Appendix \ref{App_A}), since the energy is an integral of the density of states weighted by a smearing function.

\section{Finite-temperature effects: phase stability and the Curie temperature}
\label{section:Temp_effec}  
 
 \subsection{Phase stability at finite temperatures}

For the four selected phases in the Fe$_3$Ta, Fe$_5$Ta, Fe$_5$Ta$_2$ and Fe$_6$Ta binaries, we further studied the effect of temperature and pressure on their structural stability. We have calculated the free energy by considering the lattice contribution in the quasi-harmonic approximation (QHA) and the entropy of electron system:
\begin{equation}
F(T) =  E_{0} +F_{ph}(T)-TS_{el}(T),
\label{eq:F_eph}
\end{equation}
where $E_{0}$ is the Density Functional Theory (DFT) energy at $T=0K$, $F_{ph}$ is the phonon free energy, and $S_{el}$ is the entropy of electrons, estimated by the Sommerfeld formula~\cite{Sommerfeld1928}:
\begin{equation}
S_{el}(T) =  -\frac{N(E_F)}{6}\pi^2 k_{B}^{2} T,
\label{eq:S_el}
\end{equation}
where $N(E_F)$ is the density of states (DOS) at the Fermi level.

\begin{figure}[ht!]
\centering
\includegraphics[width=0.85\columnwidth ,angle=0]{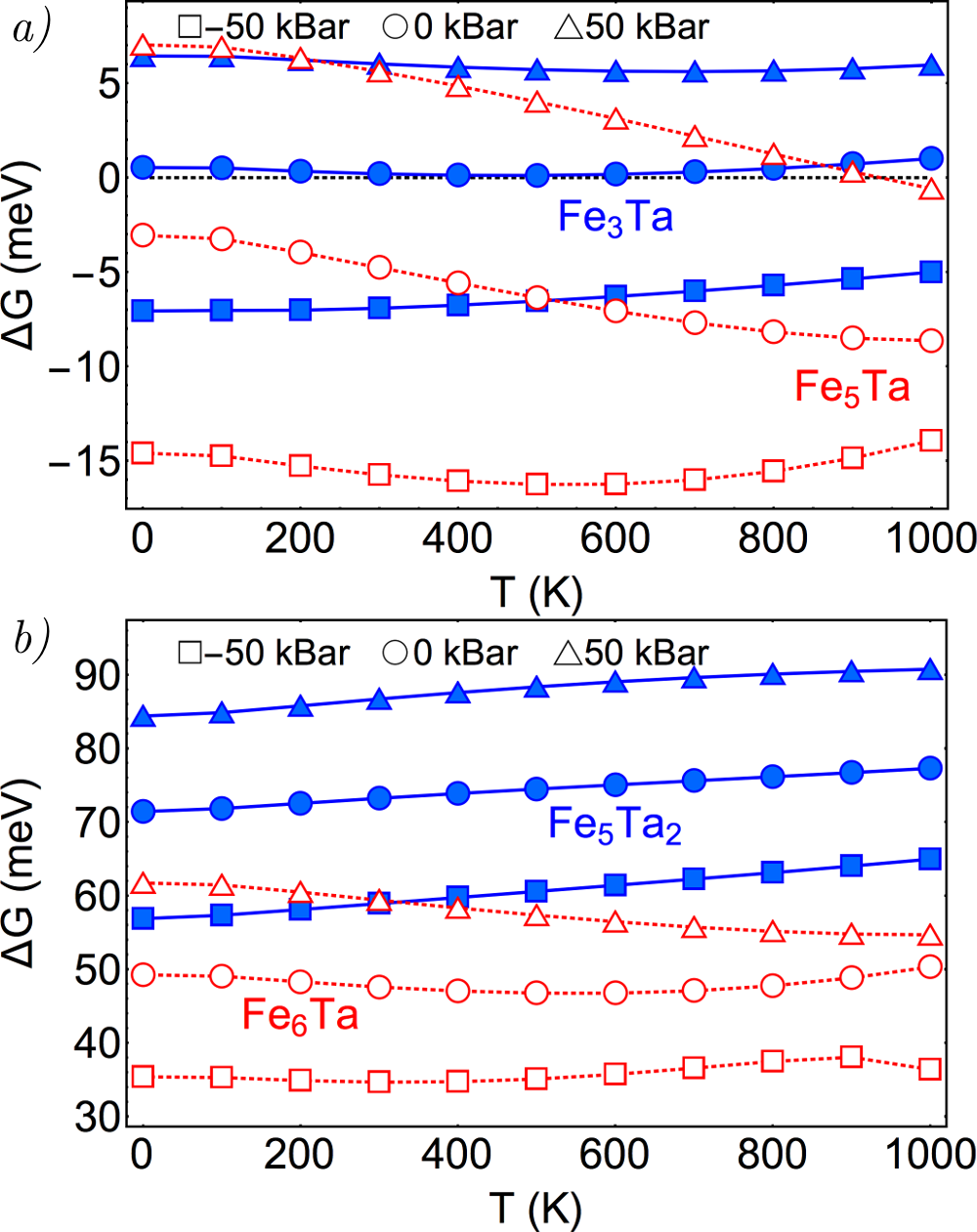}
\caption{Gibbs free energy difference $\Delta G(T,P)$  of selected structures: $a)$ Fe$_3$Ta (blue filled symbols) and Fe$_5$Ta (red open symbols); $b)$ Fe$_5$Ta$_2$ (blue filled symbols) and Fe$_6$Ta (red open symbols); relative to stable $bcc$ Fe and Fe$_2$Ta phases according to Eq.~\ref{eq:deltaG}.}
\label{fig:free_energy}
\end{figure}

Calculations were performed in the following manner. For each structure we performed DFT calculations over a set of volumes within an interval $(V_0-\Delta V, V_0+\Delta V)$ about equilibrium volume $V_0$, with $\Delta V=0.1 V_0$. Structures were relaxed at each volume and accurate total energy $E_0$ and DOS calculated. To calculate the phonon free energy $F_{ph}$ we used PHON code~\cite{ALFE20092622}, an implementation of the small displacements method to calculate phonon dispersion of a harmonic crystal. At each volume supercells were generated from relaxed structures and forces were calculated with the VASP program. For these calculations we used the PAW potentials containing the semi-core $p$ electrons, and performed accurate calculations with a cut-off energy of ENCUT~$=513.167$\,eV ($1.75$ of the default cut-off energy) and the tolerance of the electronic convergence set to EDIFF~$=10^{-7}$\,eV. We used a fully automatic k-points generation mesh with length $l=40$ to sample the reciprocal space. We have performed free energy calculations for the four above mentioned structures, as well as for the two stable reference phases: $bcc$ Fe and Fe$_2$Ta. All phases were dynamically stable within the whole considered pressure range (no imaginary frequencies). To calculate the phase stability at finite temperatures and different pressures we have estimated parameters of the equation of state (EOS) $F=F(V,T)$ at a set of temperature values within the $(0K, 1000K)$ interval by fitting at each temperature $T$ the set $\{F(V_i)\}$ of calculated free energy values to the Vinet EOS~\cite{PhysRevB.35.1945}:
\begin{eqnarray}
F(V)  =  F_0+4\frac{B_0 V_0}{(B'-1)^2} -2\frac{B_0 V_0}{(B'-1)^2}\times~~~~~~~\\
\left(5+3 B'  \left(\eta(V) -1\right)-3\eta(V)\right)\exp\left[-\frac{3}{2}\left(B'-1\right)\left(\eta(V)-1\right)\right],\nonumber
\label{eq:F_Vinet}
\end{eqnarray}
where $\eta(V)=(V/V_0)^{1/3}$ and $F_0=F_0(T)$, $V_0=V0(T)$, $B_0=B_0(T)$, and $B'=B'(T)$ are the equilibrium free energy, volume, bulk modulus and its derivative at a given $T$. Within the Vinet approximation to the Helmholtz free energy we calculated the Gibbs free energy $G(P,T)=F(V,T)+PV$ and estimated the formation energy with respect to stable $bcc$ Fe and Fe$_2$Ta as:
\begin{equation}
\Delta G=G(Fe_nTa_m)-X_n G(Fe)-Y_m G(Fe_2Ta),  
\label{eq:deltaG}
\end{equation}
where $X_n$ and $Y_m$ are appropriately chosen for each considered Fe$_n$Ta$_m$ phase. The temperature dependence of the formation energy of Fe$_3$Ta and Fe$_5$Ta, which were close to the enthalpy hull at $T=0$ (Fig.~\ref{fig:phase_ms}) is shown in Fig.~\ref{fig:free_energy}$a$  for different pressures. Calculations show that the predicted cubic Fe$_5$Ta phase should be stable at ambient pressure (data shown by open disks) over the whole considered temperature range. The promising new magnetic phase Fe$_3$Ta (shown by filled symbols) also remains very close to the stability line for the ambient pressure. At finite positive pressures (compression) both phases get energetically less stable. The effect is opposite for the negative pressure values (dilatation). This pressure effect suggests that the new phase may get energetically more stable by adding a third element, which would slightly increase the inter-atomic distances. In Fig.~\ref{fig:free_energy}$b$ we show the temperature and pressure trends for the formation energies of metastable Fe$_5$Ta$_2$ (filled symbols)  and Fe$_6$Ta (open symbols) phases with high magnetic anisotropy. Both phases remain unstable with temperature and formation energies show similar trends with pressure as in the case of Fe$_3$Ta and Fe$_5$Ta phases.

\subsection{Exchange integrals and Curie temperature}
\label{subsection:ex}  

\begin{figure}[ht!]
\centering
\includegraphics[width=\columnwidth ,angle=0]{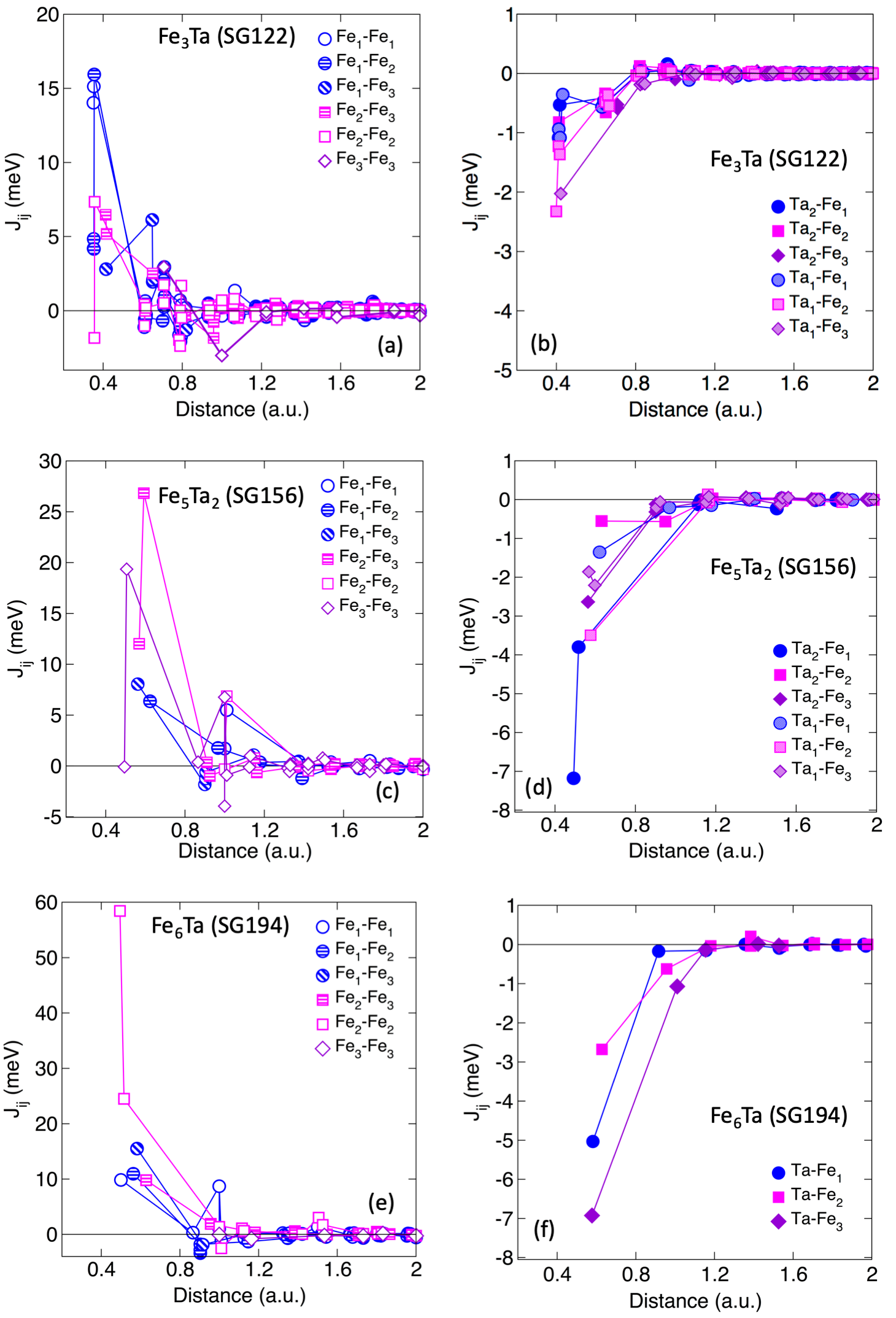}
\caption{Calculated exchange coupling constants for: (a)-(b) Fe$_3$Ta, (c)-(d) Fe$_5$Ta$_2$ and (e)-(f) Fe$_6$Ta, depending on the distance between the ions. Left: The couplings between the Fe ions in the system are shown. Right: The couplings between the magnetic Fe ions and the Ta ions (which have an induced moment antiparallel to the Fe atoms) are plotted. Note the different scale between the two graphs.}
\label{fig:exchange_Fe5Ta2}
\end{figure}

Using the optimized geometries of the new phases as the starting point, the exchange coupling constants were calculated. Using Liechtenstein's approach \cite{Jij_2} the magnetic interactions for arbitrary arrangements of magnetic moments can be calculated. The effective exchange interaction parameters were obtained using Lichtenstein \textit{et al.} method \cite{Jij_1,Jij_2}, as implemented in SPR-KKR \cite{Ebert_1996}. In this technique the energy of the system is mapped onto a classical Heisenberg model with the following Hamiltonian:
\begin{equation}
E_{ex}=-\sum_{i,j}J_{ij} \textbf{s}_i\cdot \textbf{s}_j,
\label{eq:Heis}
\end{equation}
where $J_{ij}$ are the exchange parameters and $\textbf{s}_i$ is the unit vector along the magnetic moment of atom i-th. The fast SPR-KKR core within atomic sphere approximation was used to obtain the exchange coupling constants.  After a self-consistent run needed to create the potential for the systems the $J_{ij}$ were calculated using a dense k-point mesh 34x34x29 for Fe$_5$Ta$_2$ (SG 156). The cut-off for the $J_{ij}$ couplings was chosen to be 3 lattice constants. The $J_{ij}$ values obtained from the SPR-KKR code depending on the neighbour distance are shown in the Fig. \ref{fig:exchange_Fe5Ta2}. 

Due to the complexity of the structure the Fe-Fe couplings of the Fe$_5$Ta$_2$ phase are very diverse. The contribution from Fe$_1$-Fe$_1$ (1a site, along c axis) is quite small being 5 meV for the nearest neighbors. The main contributions stem from Fe$_2$ and Fe$_3$ couplings. For these ions the next nearest neighbor couplings are between 20 and 30 meV. However, the coupling strength rapidly decreases with the distance and is almost zero after 5 neighbor shells. It should be pointed out that the majority of the coupling constants is positive which means ferromagnetic coupling. Only very tiny small antiferromagnetic contributions were observed for Fe$_5$Ta$_2$ (Fe$_3$). However, even though Ta has only an induced moment there are significant couplings between Ta and the Fe ions especially Fe$_1$. Unfortunately, the coupling is antiferromagnetic and therewith counteracts to a high T$_C$.

Finally, we estimated the T$_C$ of these novel phases by means of atomistic spin dynamics  (ASD) simulations\cite{Eriksson_book} using as inputs the calculated lattice parameters, exchange parameters and magnetic moments. For each structure, we consider a system of 15x15x15 unit cells with periodic boundary conditions, which is thermally relaxed integrating the stochastic Landau-Lifshitz-Gilbert equation. We performed this task with the software UppASD \cite{Skubic,uppasd}. The obtained values for Fe$_3$Ta, Fe$_5$Ta$_2$ and Fe$_6$Ta are T$_C$= 364 K, 724 K and 886 K, respectively, that make them suitable for PM applications. However, the performance of  Fe$_3$Ta may be highly deteriorated at room temperature since T$_C$ is not so large.
It should be noted that despite the fact that the Fe-Ta coupling parameters for Fe$_5$Ta$_2$ an Fe$_6$Ta are larger in size i.e. larger AF coupling exists, the Curie temperature for these systems are higher than for Fe$_3$Ta. This is partially related to the multiplicity of the individual couplings but also to the fact that the Fe-Fe couplings in Fe$_3$Ta are smaller.

\section{Screening of substrates for epitaxial stabilization}
\label{section:sub}

As we mentioned in the Introduction, only two stable phases have been experimentally identified for the Fe-Ta binaries so far. Our theoretical predictions seem to contradict this fact, but there may exist a possibility that observing these new phases is difficult when synthesizing the Fe-Ta alloys with traditional metallurgical methods \cite{Smallman2014}. Of course, these methods are preferable to produce a bulk PM suitable for applications. Right now, however, we would be more interested in the possibility to synthesize predicted phases and probe their magnetic properties to assess the predictive accuracy of our computational methods. One may try to stabilize some of these phases as thin films by epitaxial growth. Thus, as a final stage of our work, we performed a screening of suitable substrates that may help to stabilize these new phases.  Ding \textit{et al.} proposed two types of filters based on the unit cell topology (geometric unit cell area matching between the substrate and the target film) and strain energy density of the film in order to identify ideal substrates for epitaxial stabilization\cite{Ding}. We have applied this screening approach to 66 widely used single-crystalline substrates\cite{Ding} available in Materials Project\cite{Mat_Proj_1,Mat_Proj_2} for the films Fe$_5$Ta$_2$ (SG 156), Fe$_3$Ta (SG 122), Fe$_6$Ta (SG 194) and Fe$_5$Ta (SG 216). We performed this task using pymatgen library\cite{pymatgen}. The elastic constants needed for the calculation of the film elastic energy were obtained through AELAS code\cite{AELAS} combined with VASP using PAW method and GGA-PBE with default settings. The elastic stiffness matrix of these phases is definite positive, so that they can be considered mechanically stable crystalline structures. 
\begin{figure}[ht!]
\centering
\includegraphics[width=\columnwidth ,angle=0]{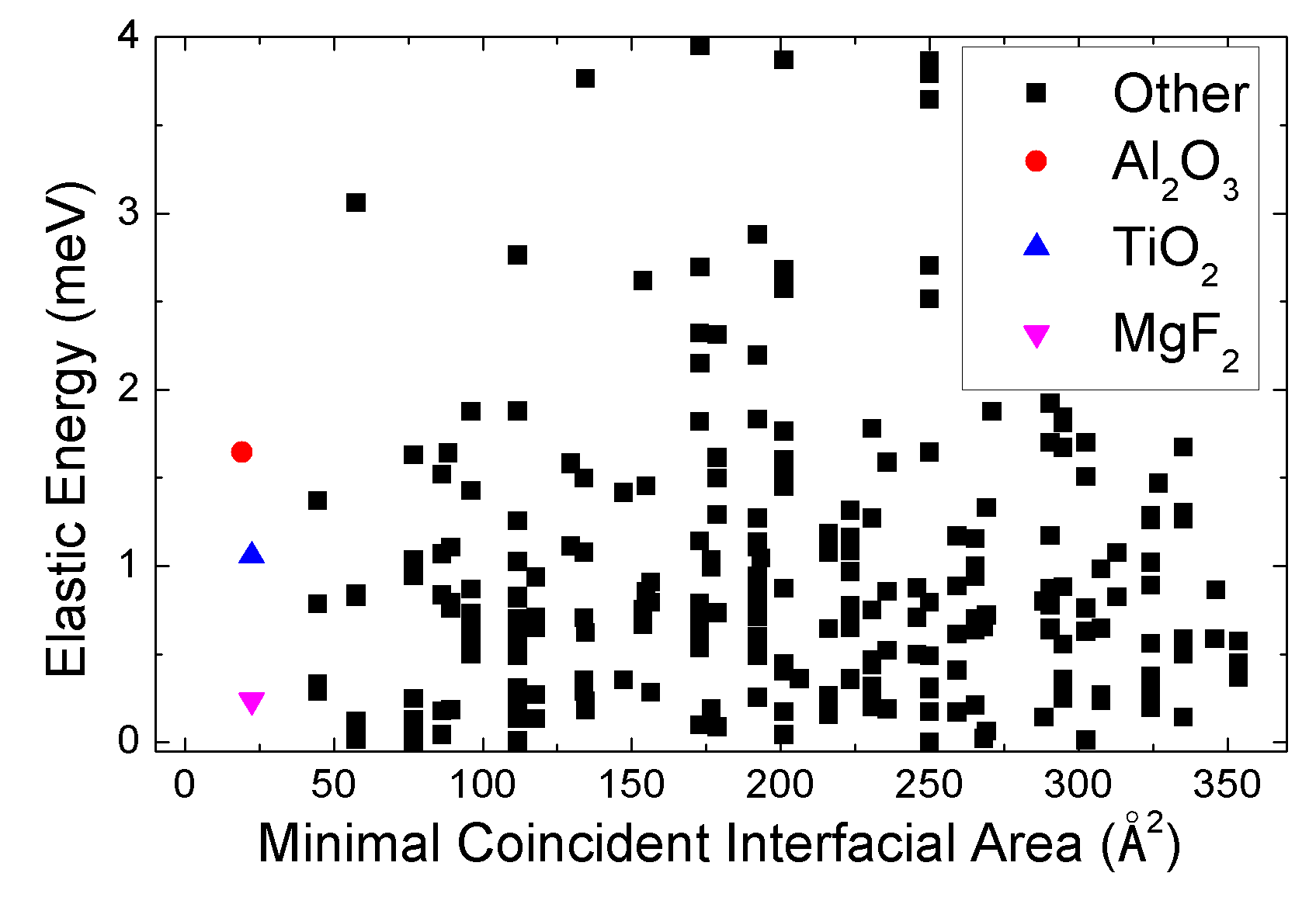}
\caption{Computational screening of substrates for the epitaxial stabilization of Fe$_5$Ta$_2$.}
\label{fig:subs_Fe5Ta2}
\end{figure}

In Fig. \ref{fig:subs_Fe5Ta2} we show the high-throughput calculation of the minimal coincident interfacial area (MCIA) and elastic energy of Fe$_5$Ta$_2$ for all considered substrates and orientations. Optimal substrates are those with both low MCIA and elastic energy. In this case, the best substrates are (001)-MgF$_2$ (001)-TiO$_2$  and (001)-Al$_2$O$_3$ with Fe$_5$Ta$_2$ film orientation (100), (100) and (001), respectively. The obtained values of MCIA and elastic energy for the substrates with the lowest value of MCIA are given in Table \ref{tab:subs} (Appendix \ref{App_C}).  Although bulk Fe$_5$Ta is not a good PM (cubic phase), its phase stability is greater than the above mentioned ones (see Fig.\ref{fig:free_energy}). Our calculations show that potential good substrates for Fe$_5$Ta could be (001)-MgF$_2$, (100)-InSb and (100)-CdTe with film orientation (100).

\section{Conclusions}
\label{section:con}

Despite of the fact that the two experimentally observed ordered Fe-Ta alloy phases do not exhibit magnetic properties suitable for PMs, our computational study suggests that new phases, with intrinsic magnetic properties appropriate for PMs, might exist within this binary system. The structure prediction study based on evolutionary algorithm revealed two new energetically stable structures for the Fe$_5$Ta and Fe$_3$Ta binaries, respectively. Fe$_3$Ta is a uniaxial phase (tetragonal symmetry) with calculated saturation magnetization $\mu_0$M$_s$=1.14 T, magnetocrystalline anisotropy K$_1$=2.17 MJ/m$^3$ and the Curie temperature $T_C$=364 K, which makes it a potentially promising phase for PMs. Calculations of MAE of various low energy metastable phases also showed the existence of structures with extraordinary high MAE in the Fe-Ta system. We identified two phases in the Fe$_5$Ta$_2$ and Fe$_6$Ta binaries with intrinsic magnetic properties comparable to SmCo$_5$ and Nd$_2$Fe$_{14}$B, respectively. For the Fe$_5$Ta$_2$ structure, for example, our calculations predict $\mu_0$M$_s$=1.00 T, K$_1$+K$_2$=16.65 MJ/m$^3$ and $T_C$=724 K.  In this phase, we found that there is a large DOS in the minority spin channel of d$_{xy}$ and d$_{x^2-y^2}$ states (d orbitals that lay on the hard plane) right at the Fermi level, especially at Wyckoff (1a) site for both Fe and Ta atoms, which may be responsible for the observed high MAE. Is is necessary, of course, to analyze more hard intermetallic Fe-based phases in order to identify possible general mechanism of high MAE. The analysis of the Gibbs free energy shows that these hard magnetic phases are energetically unstable at finite temperatures too, so this may prevent the synthesis of these phases in bulk. It might be possible, however, to synthesize these phases as thin films and we provide some possible substrates for the epitaxial growth. These new findings might constitute a step forward towards the discovery and design of novel high-performance RE-free PMs.

\section*{Acknowledgement}

This work was supported by the European Horizon 2020 Framework Programme for Research and Innovation (2014-2020) under Grant Agreement No. 686056, NOVAMAG. Authors acknowledge the European Regional Development Fund in the IT4Innovations national supercomputing center - path to exascale project, project number CZ 02.1.01/0.0/0.0/16-013/0001791 within the Operational Programme Research, Development and Education.

\section*{Appendix}

\appendix

\section{Some details of calculations of MAE}
\label{App_A}

\begin{figure}[ht!]
\centering
\includegraphics[width=0.9\columnwidth ,angle=0]{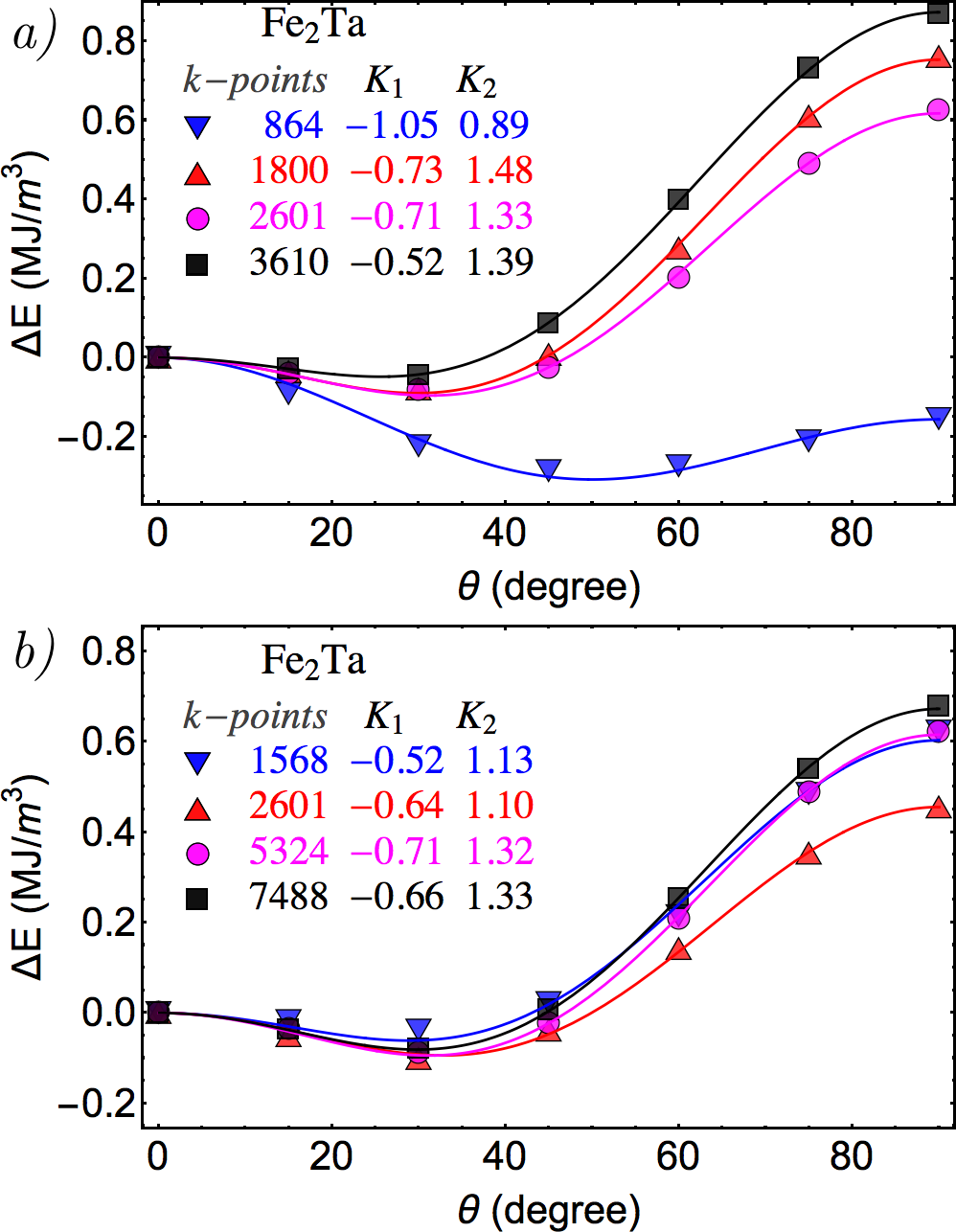}
\caption{Magneto-crystalline anisotropy energy of the ferromagnetic Laves Fe$_2$Ta phase calculated for different number of k-points in the reciprocal space: $a)$ calculations performed with PAW PBE potentials with the minimum number of valence electrons and cut-off energy of~$401.823$\,eV ($1.50$ of the default cut-off energy); $b)$ calculations performed with PAW PBE potentials with the $p$ semi-core electrons added to the valence electrons and cut-off energy of~$513.167$\,eV ($1.75$ of the default cut-off energy)}
\label{fig:mae_Fe2Ta}
\end{figure}

In this appendix, we show some tests of our MAE calculations for Laves phase Fe$_2$Ta, and theoretical predicted Fe$_3$Ta (SG 122) and Fe$_5$Ta$_2$ (SG 156). The Fe$_2$Ta phase was theoretically investigated previously~\cite{Edstrom,Kumar_2014} with contradictory results for the MAE calculations using different DFT codes. In Fig.~\ref{fig:mae_Fe2Ta} we show calculated MAE with VASP code for different k-point meshes as well as different PAW PBE potentials and cut-off energies. We can see that calculated values of the MAE are quite robust against different PAW PBE potentials and are converged for a number of k-points in the reciprocal space larger than 1500, which corresponds to the length $l=60$ in the VASP automatic k-points generation scheme. These results are in agreement with those reported by Edstr\"om~\cite{Edstrom}, who also obtained an easy cone for the MCA with $K_1=-0.27$ MJ/$m^3$ and $K_2=1.52$ MJ/$m^3$. The value of the MAE of $1.25$ MJ/$m^3$ is larger than one we had obtained (Fig.~\ref{fig:mae_Fe2Ta}) due to the larger cell volume used in Ref. \onlinecite{Edstrom}, see Table \ref{data_table}.

\begin{figure}[ht!]
\centering
\includegraphics[width=0.9\columnwidth ,angle=0]{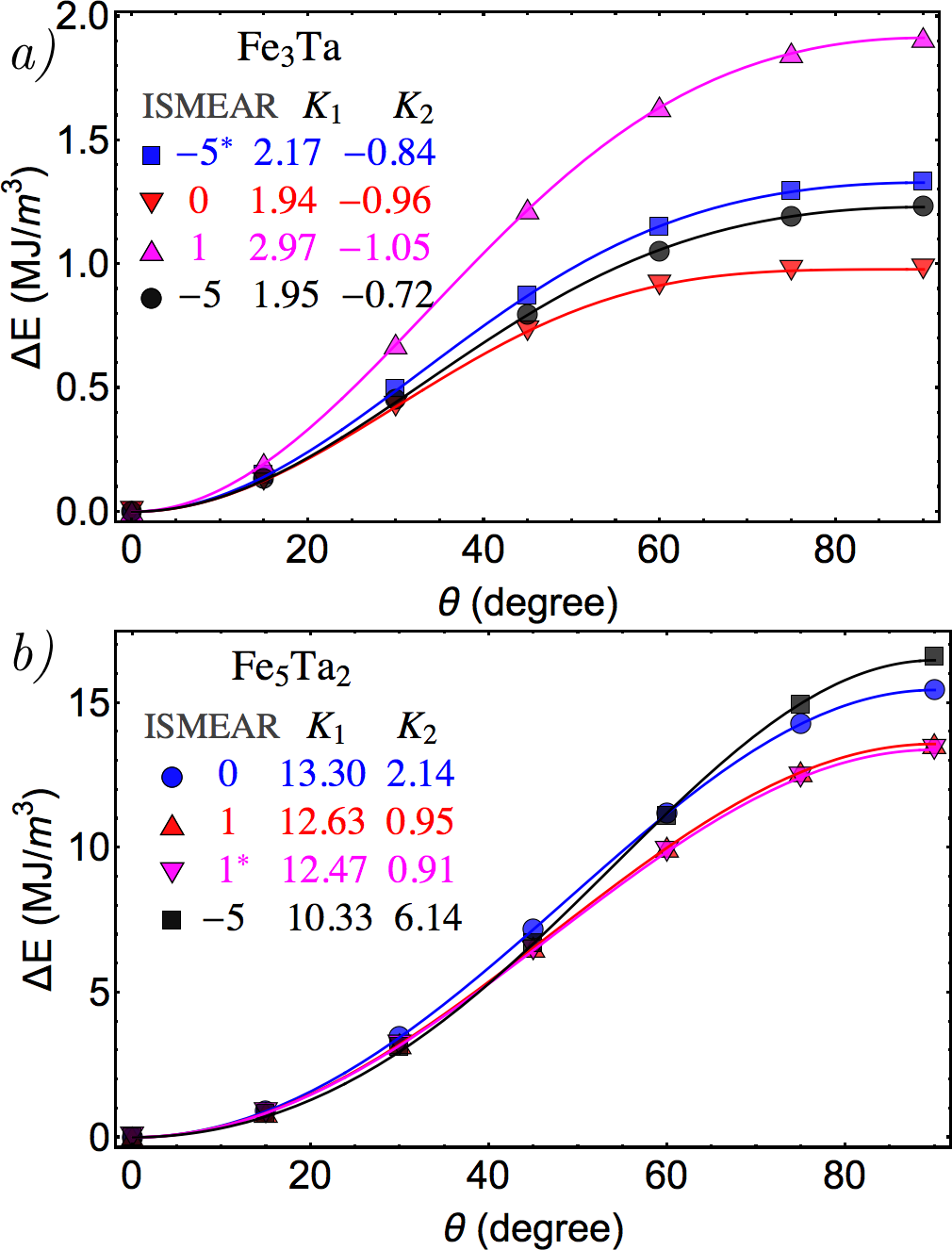}
\caption{Magneto-crystalline anisotropy energy of predicted $a)$ Fe$_3$Ta, and $b)$ Fe$_5$Ta$_2$ phases calculated for different settings for the smearing parameter ISMEAR = 0,1,-5 (Gaussian, 1st order Methfessel-Paxton and tetrahedron with Bl\"ochl corrections methods, respectively). For all of the calculations, except those marked with an asterisk, we used the PAW PBE potentials with minimum number of valence electrons, a $k-$~mesh corresponding to length $l=60$ and energy cut-off of~$401.823$\,eV ($1.50$ of the default cut-off energy). For other two calculations marked with an asterisk, we used the PAW PBE potentials with the $p$ semi-core electrons added to the valence electrons and cut-off energy ofT~$513.167$\,eV ($1.75$ of the default cut-off energy).}
\label{fig:mae_Fe5Ta2}
\end{figure}

The Fe$_3$Ta and Fe$_5$Ta$_2$ phases present a particular interest. The first one is energetically very close to the convex hull of the Fe-Ta binary diagram and, thus, could be experimentally obtained under appropriate conditions. The relatively high saturation magnetization and magnetocrystalline anisotropy make this phase a promising candidate for RE-free PM material. The second one, despite being metastable, provides an example of huge MCA for a RE-free intermetallic compound. We have performed a series of MAE calculations for different parameters which affects the energy calculations, like the type of PAW PBE potentials, energy cut-off and energy smearing. These additional calculations are shown in Fig.~\ref{fig:mae_Fe5Ta2}. We see that results of the MAE calculations are robust against the variations of these parameters.

\section{Crystallographic and magnetic data}
\label{App_B}


\begin{table}[H]
\caption{Crystallographic data and spin magnetic moment of Fe$_3$Ta (SG 122), Fe$_6$Ta (SG 194) and Fe$_5$Ta (SG 216).}
\label{data_table2}
\begin{tabular}{@{}ccccccc@{}}
Compound     & Atom & \begin{tabular}[c]{@{}c@{}}Wyckoff \\ position\end{tabular} & x     & y     & z      & \begin{tabular}[c]{@{}c@{}}$\mu_{spin}${[}001{]}\\ $(\mu_B)$\end{tabular}  \\ \midrule
Fe$_3$Ta     & Fe$_1$   & 16e                                                         & 0.876 & 0.374 & 0.436  & 1.727                                                                \\
             & Fe$_2$   & 16e                                                         & 0.626 & 0.874 & 0.061 & 1.646   \\
             & Fe$_3$   & 4b                                                          & 0   & 0   & 1/2    & 2.564                                                             \\
             & Ta$_1$   & 8d                                                          & 0.731 & 1/4   & 0.125  & -0.382                                                           \\
             & Ta$_2$   & 4a                                                          & 0   & 0   & 0    & -0.354                                                               \\ \midrule
Fe$_6$Ta     & Fe$_1$   & 6g                                                          & 1/2   & 0   & 0    & 1.921                                                              \\
             & Fe$_2$   & 4e                                                          & 0   & 0   & 0.628  & 2.436                                                             \\
             & Fe$_3$   & 2d                                                          & 1/3   & 2/3   & 3/4    & 2.305                                                              \\
             & Ta$_1$   & 2c                                                          & 1/3   & 2/3   & 1/4    & -0.890                                                                \\ \midrule
Fe$_5$Ta     & Fe$_1$   & 16e                                                         & 0.625 & 0.625 & 0.625  & 1.939                                                               \\
             & Fe$_2$   & 4c                                                          & 1/4   & 1/4   & 1/4    & 2.746                                                                                                        \\
             & Ta$_1$   & 4a                                                          & 0   & 0   & 0    & -0.438                                                              \\ \bottomrule
\end{tabular}
\end{table}


\section{Optimal substrates information for epitaxial growth}
\label{App_C}

\begin{table}[H]
\caption{Optimal substrates for epitaxial growth of phases  Fe$_5$Ta$_2$ (SG 156), Fe$_3$Ta (SG 122), Fe$_6$Ta (SG 194) and Fe$_5$Ta (SG 216).}
\label{tab:subs}
\begin{tabular}{cccccc}
\begin{tabular}[c]{@{}c@{}}Film\\ $ $\end{tabular} & \begin{tabular}[c]{@{}c@{}}Film\\ orientation\end{tabular} & \begin{tabular}[c]{@{}c@{}}Substrate\\ $ $\end{tabular} & \begin{tabular}[c]{@{}c@{}}Substrate \\ orientation\end{tabular} & \begin{tabular}[c]{@{}c@{}}MCIA\\ ($\AA^2$)\end{tabular} & \begin{tabular}[c]{@{}c@{}}Elastic energy\\ (meV)\end{tabular} \\ \hline
Fe$_5$Ta$_2$                                       & \textless{}001\textgreater{}                               & Al$_2$O$_3$                                             & \textless{}001\textgreater{}                                     & 19.24                                                    & 1.645                                                          \\
                                                   & \textless{}100\textgreater{}                               & TiO$_2$                                                 & \textless{}001\textgreater{}                                     & 22.36                                                    & 1.058                                                          \\
                                                   & \textless{}100\textgreater{}                               & MgF$_2$                                                 & \textless{}001\textgreater{}                                     & 22.36                                                    & 0.239                                                          \\ \hline
Fe$_3$Ta                                           & \textless{}001\textgreater{}                               & MgF$_2$                                                 & \textless{}001\textgreater{}                                     & 45.34                                                    & 0.793                                                          \\
                                                   & \textless{}001\textgreater{}                               & InSb                                                    & \textless{}100\textgreater{}                                     & 45.34                                                    & 0.880                                                          \\
                                                   & \textless{}001\textgreater{}                               & CdTe                                                    & \textless{}100\textgreater{}                                     & 45.34                                                    & 0.954                                                          \\ \hline
Fe$_6$Ta                                           & \textless{}100\textgreater{}                               & MgF$_2$                                                 & \textless{}100\textgreater{}                                     & 43.29                                                    & 0.213                                                          \\
                                                   & \textless{}100\textgreater{}                               & MgF$_2$                                                 & \textless{}001\textgreater{}                                     & 43.29                                                    & 0.332                                                          \\
                                                   & \textless{}100\textgreater{}                               & TiO$_2$                                                 & \textless{}001\textgreater{}                                     & 43.29                                                    & 0.041                                                          \\ \hline
Fe$_5$Ta                                           & \textless{}100\textgreater{}                               & MgF$_2$                                                 & \textless{}001\textgreater{}                                     & 44.59                                                    & 0.124                                                          \\
                                                   & \textless{}100\textgreater{}                               & InSb                                                    & \textless{}100\textgreater{}                                     & 44.59                                                    & 0.158                                                          \\
                                                   & \textless{}100\textgreater{}                               & CdTe                                                    & \textless{}100\textgreater{}                                     & 44.59                                                    & 0.189 \\\hline                                                        
\end{tabular}
\end{table}


\bibliography{mybibfile}

\begin{thebibliography}{45}
\expandafter\ifx\csname natexlab\endcsname\relax\def\natexlab#1{#1}\fi
\expandafter\ifx\csname bibnamefont\endcsname\relax
  \def\bibnamefont#1{#1}\fi
\expandafter\ifx\csname bibfnamefont\endcsname\relax
  \def\bibfnamefont#1{#1}\fi
\expandafter\ifx\csname citenamefont\endcsname\relax
  \def\citenamefont#1{#1}\fi
\expandafter\ifx\csname url\endcsname\relax
  \def\url#1{\texttt{#1}}\fi
\expandafter\ifx\csname urlprefix\endcsname\relax\def\urlprefix{URL }\fi
\providecommand{\bibinfo}[2]{#2}
\providecommand{\eprint}[2][]{\url{#2}}

\bibitem[{\citenamefont{Scheunert et~al.}(2016)\citenamefont{Scheunert,
  Heinonen, Hardeman, Lapicki, Gubbins, and Bowman}}]{Scheu}
\bibinfo{author}{\bibfnamefont{G.}~\bibnamefont{Scheunert}},
  \bibinfo{author}{\bibfnamefont{O.}~\bibnamefont{Heinonen}},
  \bibinfo{author}{\bibfnamefont{R.}~\bibnamefont{Hardeman}},
  \bibinfo{author}{\bibfnamefont{A.}~\bibnamefont{Lapicki}},
  \bibinfo{author}{\bibfnamefont{M.}~\bibnamefont{Gubbins}}, \bibnamefont{and}
  \bibinfo{author}{\bibfnamefont{R.~M.} \bibnamefont{Bowman}},
  \bibinfo{journal}{Applied Physics Reviews} \textbf{\bibinfo{volume}{3}},
  \bibinfo{pages}{011301} (\bibinfo{year}{2016}),
  \eprint{https://doi.org/10.1063/1.4941311},
  \urlprefix\url{https://doi.org/10.1063/1.4941311}.

\bibitem[{\citenamefont{Skokov and Gutfleisch}(2018)}]{SKOKOV2018289}
\bibinfo{author}{\bibfnamefont{K.}~\bibnamefont{Skokov}} \bibnamefont{and}
  \bibinfo{author}{\bibfnamefont{O.}~\bibnamefont{Gutfleisch}},
  \bibinfo{journal}{Scripta Materialia} \textbf{\bibinfo{volume}{154}},
  \bibinfo{pages}{289 } (\bibinfo{year}{2018}), ISSN \bibinfo{issn}{1359-6462},
  \urlprefix\url{http://www.sciencedirect.com/science/article/pii/S1359646218300599}.

\bibitem[{\citenamefont{Ivanov et~al.}(1973)\citenamefont{Ivanov, Solina,
  Demshina, and M.}}]{Ivanov1973}
\bibinfo{author}{\bibfnamefont{.~A.} \bibnamefont{Ivanov}},
  \bibinfo{author}{\bibfnamefont{L.~V.} \bibnamefont{Solina}},
  \bibinfo{author}{\bibfnamefont{V.~A.} \bibnamefont{Demshina}},
  \bibnamefont{and} \bibinfo{author}{\bibfnamefont{M.~L.} \bibnamefont{M.}},
  \bibinfo{journal}{Phys. Met. Metallogr.} \textbf{\bibinfo{volume}{35}},
  \bibinfo{pages}{81} (\bibinfo{year}{1973}).

\bibitem[{\citenamefont{{Phuoc} and {Ong}}(2014)}]{Phuoc1}
\bibinfo{author}{\bibfnamefont{N.~N.} \bibnamefont{{Phuoc}}} \bibnamefont{and}
  \bibinfo{author}{\bibfnamefont{C.~K.} \bibnamefont{{Ong}}},
  \bibinfo{journal}{IEEE Transactions on Magnetics}
  \textbf{\bibinfo{volume}{50}}, \bibinfo{pages}{1} (\bibinfo{year}{2014}),
  ISSN \bibinfo{issn}{0018-9464}.

\bibitem[{\citenamefont{Phuoc et~al.}(2013)\citenamefont{Phuoc, Chapon, Acher,
  and Ong}}]{Phuoc2}
\bibinfo{author}{\bibfnamefont{N.~N.} \bibnamefont{Phuoc}},
  \bibinfo{author}{\bibfnamefont{P.}~\bibnamefont{Chapon}},
  \bibinfo{author}{\bibfnamefont{O.}~\bibnamefont{Acher}}, \bibnamefont{and}
  \bibinfo{author}{\bibfnamefont{C.~K.} \bibnamefont{Ong}},
  \bibinfo{journal}{Journal of Applied Physics} \textbf{\bibinfo{volume}{114}},
  \bibinfo{pages}{153903} (\bibinfo{year}{2013}),
  \eprint{https://doi.org/10.1063/1.4825225},
  \urlprefix\url{https://doi.org/10.1063/1.4825225}.

\bibitem[{\citenamefont{Taskaev et~al.}(2015)\citenamefont{Taskaev, Skokov,
  Khovaylo, Gunderov, and Karpenkov}}]{Taskaev}
\bibinfo{author}{\bibfnamefont{S.}~\bibnamefont{Taskaev}},
  \bibinfo{author}{\bibfnamefont{K.}~\bibnamefont{Skokov}},
  \bibinfo{author}{\bibfnamefont{V.}~\bibnamefont{Khovaylo}},
  \bibinfo{author}{\bibfnamefont{D.}~\bibnamefont{Gunderov}}, \bibnamefont{and}
  \bibinfo{author}{\bibfnamefont{D.}~\bibnamefont{Karpenkov}},
  \bibinfo{journal}{Physics Procedia} \textbf{\bibinfo{volume}{75}},
  \bibinfo{pages}{1404 } (\bibinfo{year}{2015}), ISSN
  \bibinfo{issn}{1875-3892}, \bibinfo{note}{20th International Conference on
  Magnetism, ICM 2015},
  \urlprefix\url{http://www.sciencedirect.com/science/article/pii/S1875389215017976}.

\bibitem[{\citenamefont{Lamichhane et~al.}(2019)\citenamefont{Lamichhane,
  Onyszczak, Palasyuk, Sharikadze, Kim, Lin, Kramer, McCallum, Wysocki, Nguyen
  et~al.}}]{Lami2019}
\bibinfo{author}{\bibfnamefont{T.~N.} \bibnamefont{Lamichhane}},
  \bibinfo{author}{\bibfnamefont{M.~T.} \bibnamefont{Onyszczak}},
  \bibinfo{author}{\bibfnamefont{O.}~\bibnamefont{Palasyuk}},
  \bibinfo{author}{\bibfnamefont{S.}~\bibnamefont{Sharikadze}},
  \bibinfo{author}{\bibfnamefont{T.-H.} \bibnamefont{Kim}},
  \bibinfo{author}{\bibfnamefont{Q.}~\bibnamefont{Lin}},
  \bibinfo{author}{\bibfnamefont{M.~J.} \bibnamefont{Kramer}},
  \bibinfo{author}{\bibfnamefont{R.}~\bibnamefont{McCallum}},
  \bibinfo{author}{\bibfnamefont{A.~L.} \bibnamefont{Wysocki}},
  \bibinfo{author}{\bibfnamefont{M.~C.} \bibnamefont{Nguyen}},
  \bibnamefont{et~al.}, \bibinfo{journal}{Phys. Rev. Applied}
  \textbf{\bibinfo{volume}{11}}, \bibinfo{pages}{014052}
  (\bibinfo{year}{2019}),
  \urlprefix\url{https://link.aps.org/doi/10.1103/PhysRevApplied.11.014052}.

\bibitem[{\citenamefont{Villars et~al.}(2014)\citenamefont{Villars, Cenzual,
  and Gladyshevskii}}]{Villars2014}
\bibinfo{author}{\bibfnamefont{P.}~\bibnamefont{Villars}},
  \bibinfo{author}{\bibfnamefont{K.}~\bibnamefont{Cenzual}}, \bibnamefont{and}
  \bibinfo{author}{\bibfnamefont{R.}~\bibnamefont{Gladyshevskii}},
  \emph{\bibinfo{title}{Handbook of Inorganic Substances 2014}}
  (\bibinfo{publisher}{De {G}ruyter}, \bibinfo{year}{2014}).

\bibitem[{\citenamefont{Shinagawa et~al.}(2014)\citenamefont{Shinagawa, Chinen,
  Omori, Oikawa, Ohnuma, Ishida, and Kainuma}}]{SHINAGAWA201487}
\bibinfo{author}{\bibfnamefont{K.}~\bibnamefont{Shinagawa}},
  \bibinfo{author}{\bibfnamefont{H.}~\bibnamefont{Chinen}},
  \bibinfo{author}{\bibfnamefont{T.}~\bibnamefont{Omori}},
  \bibinfo{author}{\bibfnamefont{K.}~\bibnamefont{Oikawa}},
  \bibinfo{author}{\bibfnamefont{I.}~\bibnamefont{Ohnuma}},
  \bibinfo{author}{\bibfnamefont{K.}~\bibnamefont{Ishida}}, \bibnamefont{and}
  \bibinfo{author}{\bibfnamefont{R.}~\bibnamefont{Kainuma}},
  \bibinfo{journal}{Intermetallics} \textbf{\bibinfo{volume}{49}},
  \bibinfo{pages}{87 } (\bibinfo{year}{2014}), ISSN \bibinfo{issn}{0966-9795},
  \urlprefix\url{http://www.sciencedirect.com/science/article/pii/S096697951400034X}.

\bibitem[{\citenamefont{Kai et~al.}(1970)\citenamefont{Kai, Nakamichi, and
  Yamamoto}}]{Kai}
\bibinfo{author}{\bibfnamefont{K.}~\bibnamefont{Kai}},
  \bibinfo{author}{\bibfnamefont{T.}~\bibnamefont{Nakamichi}},
  \bibnamefont{and} \bibinfo{author}{\bibfnamefont{M.}~\bibnamefont{Yamamoto}},
  \bibinfo{journal}{Journal of the Physical Society of Japan}
  \textbf{\bibinfo{volume}{29}}, \bibinfo{pages}{1094} (\bibinfo{year}{1970}),
  \eprint{https://doi.org/10.1143/JPSJ.29.1094},
  \urlprefix\url{https://doi.org/10.1143/JPSJ.29.1094}.

\bibitem[{\citenamefont{Edstr\"om}(2017)}]{Edstrom}
\bibinfo{author}{\bibfnamefont{A.}~\bibnamefont{Edstr\"om}},
  \bibinfo{journal}{Phys. Rev. B} \textbf{\bibinfo{volume}{96}},
  \bibinfo{pages}{064422} (\bibinfo{year}{2017}),
  \urlprefix\url{https://link.aps.org/doi/10.1103/PhysRevB.96.064422}.

\bibitem[{\citenamefont{Gabay and Hadjipanayis}(2019)}]{Fe2Ta_George}
\bibinfo{author}{\bibfnamefont{A.~M.} \bibnamefont{Gabay}} \bibnamefont{and}
  \bibinfo{author}{\bibfnamefont{G.~C.} \bibnamefont{Hadjipanayis}},
  \bibinfo{journal}{AIP Advances} \textbf{\bibinfo{volume}{9}},
  \bibinfo{pages}{035143} (\bibinfo{year}{2019}),
  \eprint{https://doi.org/10.1063/1.5079727},
  \urlprefix\url{https://doi.org/10.1063/1.5079727}.

\bibitem[{\citenamefont{Ahmed et~al.}(1983)\citenamefont{Ahmed, Hallam, and
  Read}}]{Ahmed}
\bibinfo{author}{\bibfnamefont{M.}~\bibnamefont{Ahmed}},
  \bibinfo{author}{\bibfnamefont{G.}~\bibnamefont{Hallam}}, \bibnamefont{and}
  \bibinfo{author}{\bibfnamefont{D.}~\bibnamefont{Read}},
  \bibinfo{journal}{Journal of Magnetism and Magnetic Materials}
  \textbf{\bibinfo{volume}{37}}, \bibinfo{pages}{101 } (\bibinfo{year}{1983}),
  ISSN \bibinfo{issn}{0304-8853},
  \urlprefix\url{http://www.sciencedirect.com/science/article/pii/030488538390358X}.

\bibitem[{\citenamefont{Shaji et~al.}(2019)\citenamefont{Shaji, Mucha,
  Majumdar, Binek, Kebede, and Kumar}}]{SHAJI}
\bibinfo{author}{\bibfnamefont{S.}~\bibnamefont{Shaji}},
  \bibinfo{author}{\bibfnamefont{N.~R.} \bibnamefont{Mucha}},
  \bibinfo{author}{\bibfnamefont{A.}~\bibnamefont{Majumdar}},
  \bibinfo{author}{\bibfnamefont{C.}~\bibnamefont{Binek}},
  \bibinfo{author}{\bibfnamefont{A.}~\bibnamefont{Kebede}}, \bibnamefont{and}
  \bibinfo{author}{\bibfnamefont{D.}~\bibnamefont{Kumar}},
  \bibinfo{journal}{Journal of Magnetism and Magnetic Materials}
  \textbf{\bibinfo{volume}{489}}, \bibinfo{pages}{165446}
  (\bibinfo{year}{2019}), ISSN \bibinfo{issn}{0304-8853},
  \urlprefix\url{http://www.sciencedirect.com/science/article/pii/S0304885318328907}.

\bibitem[{Afl()}]{Aflow_1}
\emph{\bibinfo{title}{Aflowlib}},
  \bibinfo{howpublished}{\url{{http://aflowlib.org/}}}.

\bibitem[{\citenamefont{Lyakhov et~al.}(2013)\citenamefont{Lyakhov, Oganov,
  Stokes, and Zhu}}]{uspex}
\bibinfo{author}{\bibfnamefont{A.~O.} \bibnamefont{Lyakhov}},
  \bibinfo{author}{\bibfnamefont{A.~R.} \bibnamefont{Oganov}},
  \bibinfo{author}{\bibfnamefont{H.~T.} \bibnamefont{Stokes}},
  \bibnamefont{and} \bibinfo{author}{\bibfnamefont{Q.}~\bibnamefont{Zhu}},
  \bibinfo{journal}{Computer Physics Communications}
  \textbf{\bibinfo{volume}{184}}, \bibinfo{pages}{1172 }
  (\bibinfo{year}{2013}), ISSN \bibinfo{issn}{0010-4655},
  \urlprefix\url{http://www.sciencedirect.com/science/article/pii/S0010465512004055}.

\bibitem[{usp()}]{uspex_web}
\bibinfo{howpublished}{\url{https://uspex-team.org/en}}.

\bibitem[{\citenamefont{Kresse and Hafner}(1993)}]{vasp_1}
\bibinfo{author}{\bibfnamefont{G.}~\bibnamefont{Kresse}} \bibnamefont{and}
  \bibinfo{author}{\bibfnamefont{J.}~\bibnamefont{Hafner}},
  \bibinfo{journal}{Phys. Rev. B} \textbf{\bibinfo{volume}{47}},
  \bibinfo{pages}{558} (\bibinfo{year}{1993}),
  \urlprefix\url{https://link.aps.org/doi/10.1103/PhysRevB.47.558}.

\bibitem[{\citenamefont{Kresse and Furthmüller}(1996)}]{vasp_2}
\bibinfo{author}{\bibfnamefont{G.}~\bibnamefont{Kresse}} \bibnamefont{and}
  \bibinfo{author}{\bibfnamefont{J.}~\bibnamefont{Furthmüller}},
  \bibinfo{journal}{Computational Materials Science}
  \textbf{\bibinfo{volume}{6}}, \bibinfo{pages}{15 } (\bibinfo{year}{1996}),
  ISSN \bibinfo{issn}{0927-0256},
  \urlprefix\url{http://www.sciencedirect.com/science/article/pii/0927025696000080}.

\bibitem[{\citenamefont{Kresse and Furthm\"uller}(1996)}]{vasp_3}
\bibinfo{author}{\bibfnamefont{G.}~\bibnamefont{Kresse}} \bibnamefont{and}
  \bibinfo{author}{\bibfnamefont{J.}~\bibnamefont{Furthm\"uller}},
  \bibinfo{journal}{Phys. Rev. B} \textbf{\bibinfo{volume}{54}},
  \bibinfo{pages}{11169} (\bibinfo{year}{1996}),
  \urlprefix\url{https://link.aps.org/doi/10.1103/PhysRevB.54.11169}.

\bibitem[{\citenamefont{Arapan et~al.}(2018)\citenamefont{Arapan, Nieves, and
  Cuesta-L{\'o}pez}}]{Arapan_AGA_1}
\bibinfo{author}{\bibfnamefont{S.}~\bibnamefont{Arapan}},
  \bibinfo{author}{\bibfnamefont{P.}~\bibnamefont{Nieves}}, \bibnamefont{and}
  \bibinfo{author}{\bibfnamefont{S.}~\bibnamefont{Cuesta-L{\'o}pez}},
  \bibinfo{journal}{J. Appl. Phys.} \textbf{\bibinfo{volume}{123}},
  \bibinfo{pages}{083904} (\bibinfo{year}{2018}).

\bibitem[{\citenamefont{Bl\"ochl}(1994)}]{PAW}
\bibinfo{author}{\bibfnamefont{P.~E.} \bibnamefont{Bl\"ochl}},
  \bibinfo{journal}{Phys. Rev. B} \textbf{\bibinfo{volume}{50}},
  \bibinfo{pages}{17953} (\bibinfo{year}{1994}),
  \urlprefix\url{http://link.aps.org/doi/10.1103/PhysRevB.50.17953}.

\bibitem[{\citenamefont{Perdew et~al.}(1996)\citenamefont{Perdew, Burke, and
  Ernzerhof}}]{PBE}
\bibinfo{author}{\bibfnamefont{J.~P.} \bibnamefont{Perdew}},
  \bibinfo{author}{\bibfnamefont{K.}~\bibnamefont{Burke}}, \bibnamefont{and}
  \bibinfo{author}{\bibfnamefont{M.}~\bibnamefont{Ernzerhof}},
  \bibinfo{journal}{Phys. Rev. Lett.} \textbf{\bibinfo{volume}{77}},
  \bibinfo{pages}{3865} (\bibinfo{year}{1996}),
  \urlprefix\url{http://link.aps.org/doi/10.1103/PhysRevLett.77.3865}.

\bibitem[{\citenamefont{Curtarolo et~al.}(2012)\citenamefont{Curtarolo,
  Setyawan, Wang, Xue, Yang, Taylor, Nelson, Hart, Sanvito, Buongiorno-Nardelli
  et~al.}}]{Aflow_2}
\bibinfo{author}{\bibfnamefont{S.}~\bibnamefont{Curtarolo}},
  \bibinfo{author}{\bibfnamefont{W.}~\bibnamefont{Setyawan}},
  \bibinfo{author}{\bibfnamefont{S.}~\bibnamefont{Wang}},
  \bibinfo{author}{\bibfnamefont{J.}~\bibnamefont{Xue}},
  \bibinfo{author}{\bibfnamefont{K.}~\bibnamefont{Yang}},
  \bibinfo{author}{\bibfnamefont{R.~H.} \bibnamefont{Taylor}},
  \bibinfo{author}{\bibfnamefont{L.~J.} \bibnamefont{Nelson}},
  \bibinfo{author}{\bibfnamefont{G.~L.} \bibnamefont{Hart}},
  \bibinfo{author}{\bibfnamefont{S.}~\bibnamefont{Sanvito}},
  \bibinfo{author}{\bibfnamefont{M.}~\bibnamefont{Buongiorno-Nardelli}},
  \bibnamefont{et~al.}, \bibinfo{journal}{Comp. Mater. Sci.}
  \textbf{\bibinfo{volume}{58}}, \bibinfo{pages}{227} (\bibinfo{year}{2012}).

\bibitem[{nov()}]{novamag}
\bibinfo{howpublished}{\url{http://crono.ubu.es/novamag/}}.

\bibitem[{\citenamefont{Nieves et~al.}(2019)\citenamefont{Nieves, Arapan,
  Maudes-Raedo, Marticorena-S{\'a}nchez, Br{\'i}o, Kovacs, Echevarria-Bonet,
  Salazar, Weischenberg, Zhang et~al.}}]{Nieves_2019}
\bibinfo{author}{\bibfnamefont{P.}~\bibnamefont{Nieves}},
  \bibinfo{author}{\bibfnamefont{S.}~\bibnamefont{Arapan}},
  \bibinfo{author}{\bibfnamefont{J.}~\bibnamefont{Maudes-Raedo}},
  \bibinfo{author}{\bibfnamefont{R.}~\bibnamefont{Marticorena-S{\'a}nchez}},
  \bibinfo{author}{\bibfnamefont{N.~D.} \bibnamefont{Br{\'i}o}},
  \bibinfo{author}{\bibfnamefont{A.}~\bibnamefont{Kovacs}},
  \bibinfo{author}{\bibfnamefont{C.}~\bibnamefont{Echevarria-Bonet}},
  \bibinfo{author}{\bibfnamefont{D.}~\bibnamefont{Salazar}},
  \bibinfo{author}{\bibfnamefont{J.}~\bibnamefont{Weischenberg}},
  \bibinfo{author}{\bibfnamefont{H.}~\bibnamefont{Zhang}},
  \bibnamefont{et~al.}, \bibinfo{journal}{Computational Materials Science}
  \textbf{\bibinfo{volume}{168}}, \bibinfo{pages}{188 } (\bibinfo{year}{2019}),
  ISSN \bibinfo{issn}{0927-0256},
  \urlprefix\url{http://www.sciencedirect.com/science/article/pii/S0927025619303489}.

\bibitem[{\citenamefont{Coey}(2010)}]{Coeybook}
\bibinfo{author}{\bibfnamefont{J.~M.~D.} \bibnamefont{Coey}},
  \emph{\bibinfo{title}{Magnetism and Magnetic Materials}}
  (\bibinfo{publisher}{Cambridge University Press}, \bibinfo{year}{2010}).

\bibitem[{\citenamefont{Antropov et~al.}(2014)\citenamefont{Antropov, Ke, and
  {\AA}berg}}]{antropov2014}
\bibinfo{author}{\bibfnamefont{V.}~\bibnamefont{Antropov}},
  \bibinfo{author}{\bibfnamefont{L.}~\bibnamefont{Ke}}, \bibnamefont{and}
  \bibinfo{author}{\bibfnamefont{D.}~\bibnamefont{{\AA}berg}},
  \bibinfo{journal}{Solid State Communications} \textbf{\bibinfo{volume}{194}},
  \bibinfo{pages}{35 } (\bibinfo{year}{2014}), ISSN \bibinfo{issn}{0038-1098},
  \urlprefix\url{http://www.sciencedirect.com/science/article/pii/S0038109814002476}.

\bibitem[{\citenamefont{Liu et~al.}(2019)\citenamefont{Liu, Legut, Zhang, Wang,
  Fan, and Zhang}}]{Liu2019}
\bibinfo{author}{\bibfnamefont{X.}~\bibnamefont{Liu}},
  \bibinfo{author}{\bibfnamefont{D.}~\bibnamefont{Legut}},
  \bibinfo{author}{\bibfnamefont{R.}~\bibnamefont{Zhang}},
  \bibinfo{author}{\bibfnamefont{T.}~\bibnamefont{Wang}},
  \bibinfo{author}{\bibfnamefont{Y.}~\bibnamefont{Fan}}, \bibnamefont{and}
  \bibinfo{author}{\bibfnamefont{Q.}~\bibnamefont{Zhang}},
  \bibinfo{journal}{Phys. Rev. B} \textbf{\bibinfo{volume}{100}},
  \bibinfo{pages}{054438} (\bibinfo{year}{2019}),
  \urlprefix\url{https://link.aps.org/doi/10.1103/PhysRevB.100.054438}.

\bibitem[{\citenamefont{Sommerfeld}(1928)}]{Sommerfeld1928}
\bibinfo{author}{\bibfnamefont{A.}~\bibnamefont{Sommerfeld}},
  \bibinfo{journal}{Zeitschrift f{\"u}r Physik} \textbf{\bibinfo{volume}{47}},
  \bibinfo{pages}{1} (\bibinfo{year}{1928}), ISSN \bibinfo{issn}{0044-3328},
  \urlprefix\url{https://doi.org/10.1007/BF01391052}.

\bibitem[{\citenamefont{Alf√®}(2009)}]{ALFE20092622}
\bibinfo{author}{\bibfnamefont{D.}~\bibnamefont{Alf√®}},
  \bibinfo{journal}{Computer Physics Communications}
  \textbf{\bibinfo{volume}{180}}, \bibinfo{pages}{2622 }
  (\bibinfo{year}{2009}), ISSN \bibinfo{issn}{0010-4655}, \bibinfo{note}{40
  YEARS OF CPC: A celebratory issue focused on quality software for high
  performance, grid and novel computing architectures},
  \urlprefix\url{http://www.sciencedirect.com/science/article/pii/S0010465509001064}.

\bibitem[{\citenamefont{Vinet et~al.}(1987)\citenamefont{Vinet, Smith,
  Ferrante, and Rose}}]{PhysRevB.35.1945}
\bibinfo{author}{\bibfnamefont{P.}~\bibnamefont{Vinet}},
  \bibinfo{author}{\bibfnamefont{J.~R.} \bibnamefont{Smith}},
  \bibinfo{author}{\bibfnamefont{J.}~\bibnamefont{Ferrante}}, \bibnamefont{and}
  \bibinfo{author}{\bibfnamefont{J.~H.} \bibnamefont{Rose}},
  \bibinfo{journal}{Phys. Rev. B} \textbf{\bibinfo{volume}{35}},
  \bibinfo{pages}{1945} (\bibinfo{year}{1987}),
  \urlprefix\url{https://link.aps.org/doi/10.1103/PhysRevB.35.1945}.

\bibitem[{\citenamefont{Liechtenstein et~al.}(1987)\citenamefont{Liechtenstein,
  Katsnelson, Antropov, and Gubanov}}]{Jij_2}
\bibinfo{author}{\bibfnamefont{A.}~\bibnamefont{Liechtenstein}},
  \bibinfo{author}{\bibfnamefont{M.}~\bibnamefont{Katsnelson}},
  \bibinfo{author}{\bibfnamefont{V.}~\bibnamefont{Antropov}}, \bibnamefont{and}
  \bibinfo{author}{\bibfnamefont{V.}~\bibnamefont{Gubanov}},
  \bibinfo{journal}{Journal of Magnetism and Magnetic Materials}
  \textbf{\bibinfo{volume}{67}}, \bibinfo{pages}{65 } (\bibinfo{year}{1987}),
  ISSN \bibinfo{issn}{0304--8853},
  \urlprefix\url{http://www.sciencedirect.com/science/article/pii/0304885387907219}.

\bibitem[{\citenamefont{Katsnelson and Lichtenstein}(2000)}]{Jij_1}
\bibinfo{author}{\bibfnamefont{M.~I.} \bibnamefont{Katsnelson}}
  \bibnamefont{and} \bibinfo{author}{\bibfnamefont{A.~I.}
  \bibnamefont{Lichtenstein}}, \bibinfo{journal}{Phys. Rev. B}
  \textbf{\bibinfo{volume}{61}}, \bibinfo{pages}{8906} (\bibinfo{year}{2000}),
  \urlprefix\url{https://link.aps.org/doi/10.1103/PhysRevB.61.8906}.

\bibitem[{\citenamefont{Ebert}(1996)}]{Ebert_1996}
\bibinfo{author}{\bibfnamefont{H.}~\bibnamefont{Ebert}},
  \bibinfo{journal}{Reports on Progress in Physics}
  \textbf{\bibinfo{volume}{59}}, \bibinfo{pages}{1665} (\bibinfo{year}{1996}),
  \urlprefix\url{https://doi.org/10.1088%2F0034-4885%2F59%2F12%2F003}.

\bibitem[{\citenamefont{Eriksson et~al.}(2017)\citenamefont{Eriksson, Bergman,
  Bergqvist, and Hellsvik}}]{Eriksson_book}
\bibinfo{author}{\bibfnamefont{O.}~\bibnamefont{Eriksson}},
  \bibinfo{author}{\bibfnamefont{A.}~\bibnamefont{Bergman}},
  \bibinfo{author}{\bibfnamefont{L.}~\bibnamefont{Bergqvist}},
  \bibnamefont{and} \bibinfo{author}{\bibfnamefont{J.}~\bibnamefont{Hellsvik}},
  \emph{\bibinfo{title}{Atomistic Spin Dynamics Foundations and Applications}}
  (\bibinfo{publisher}{Oxford University Press}, \bibinfo{year}{2017}).

\bibitem[{\citenamefont{Skubic et~al.}(2008)\citenamefont{Skubic, Hellsvik,
  Nordström, and Eriksson}}]{Skubic}
\bibinfo{author}{\bibfnamefont{B.}~\bibnamefont{Skubic}},
  \bibinfo{author}{\bibfnamefont{J.}~\bibnamefont{Hellsvik}},
  \bibinfo{author}{\bibfnamefont{L.}~\bibnamefont{Nordström}},
  \bibnamefont{and} \bibinfo{author}{\bibfnamefont{O.}~\bibnamefont{Eriksson}},
  \bibinfo{journal}{Journal of Physics: Condensed Matter}
  \textbf{\bibinfo{volume}{20}}, \bibinfo{pages}{315203}
  (\bibinfo{year}{2008}),
  \urlprefix\url{https://doi.org/10.1088%2F0953-8984%2F20%2F31%2F315203}.

\bibitem[{upp()}]{uppasd}
\bibinfo{howpublished}{\url{http://www.physics.uu.se/research/materials-theory/ongoingresearch/uppasd/}}.

\bibitem[{\citenamefont{Smallman and Ngan}(2014)}]{Smallman2014}
\bibinfo{editor}{\bibfnamefont{R.}~\bibnamefont{Smallman}} \bibnamefont{and}
  \bibinfo{editor}{\bibfnamefont{A.}~\bibnamefont{Ngan}}, eds.,
  \emph{\bibinfo{title}{Modern Physical Metallurgy (Eighth Edition)}}
  (\bibinfo{publisher}{Butterworth-Heinemann}, \bibinfo{address}{Oxford},
  \bibinfo{year}{2014}), \bibinfo{edition}{eighth edition} ed., ISBN
  \bibinfo{isbn}{978-0-08-098204-5},
  \urlprefix\url{http://www.sciencedirect.com/science/article/pii/B9780080982045000171}.

\bibitem[{\citenamefont{Ding et~al.}(2016)\citenamefont{Ding, Dwaraknath,
  Garten, Ndione, Ginley, and Persson}}]{Ding}
\bibinfo{author}{\bibfnamefont{H.}~\bibnamefont{Ding}},
  \bibinfo{author}{\bibfnamefont{S.~S.} \bibnamefont{Dwaraknath}},
  \bibinfo{author}{\bibfnamefont{L.}~\bibnamefont{Garten}},
  \bibinfo{author}{\bibfnamefont{P.}~\bibnamefont{Ndione}},
  \bibinfo{author}{\bibfnamefont{D.}~\bibnamefont{Ginley}}, \bibnamefont{and}
  \bibinfo{author}{\bibfnamefont{K.~A.} \bibnamefont{Persson}},
  \bibinfo{journal}{ACS Applied Materials $\&$ Interfaces}
  \textbf{\bibinfo{volume}{8}}, \bibinfo{pages}{13086} (\bibinfo{year}{2016}).

\bibitem[{\citenamefont{Jain et~al.}(2013)\citenamefont{Jain, Ong, Hautier,
  Chen, Richards, Dacek, Cholia, Gunter, Skinner, Ceder et~al.}}]{Mat_Proj_1}
\bibinfo{author}{\bibfnamefont{A.}~\bibnamefont{Jain}},
  \bibinfo{author}{\bibfnamefont{S.}~\bibnamefont{Ong}},
  \bibinfo{author}{\bibfnamefont{G.}~\bibnamefont{Hautier}},
  \bibinfo{author}{\bibfnamefont{W.}~\bibnamefont{Chen}},
  \bibinfo{author}{\bibfnamefont{W.}~\bibnamefont{Richards}},
  \bibinfo{author}{\bibfnamefont{S.}~\bibnamefont{Dacek}},
  \bibinfo{author}{\bibfnamefont{S.}~\bibnamefont{Cholia}},
  \bibinfo{author}{\bibfnamefont{D.}~\bibnamefont{Gunter}},
  \bibinfo{author}{\bibfnamefont{D.}~\bibnamefont{Skinner}},
  \bibinfo{author}{\bibfnamefont{G.}~\bibnamefont{Ceder}},
  \bibnamefont{et~al.}, \bibinfo{journal}{APL Materials}
  \textbf{\bibinfo{volume}{1}}, \bibinfo{pages}{011002} (\bibinfo{year}{2013}).

\bibitem[{Mat()}]{Mat_Proj_2}
\emph{\bibinfo{title}{The materials project}},
  \bibinfo{howpublished}{\url{https://materialsproject.org/}}.

\bibitem[{\citenamefont{Ong et~al.}(2013)\citenamefont{Ong, Richards, Jain,
  Hautier, Kocher, Cholia, Gunter, Chevrier, Persson, and Ceder}}]{pymatgen}
\bibinfo{author}{\bibfnamefont{S.~P.} \bibnamefont{Ong}},
  \bibinfo{author}{\bibfnamefont{W.~D.} \bibnamefont{Richards}},
  \bibinfo{author}{\bibfnamefont{A.}~\bibnamefont{Jain}},
  \bibinfo{author}{\bibfnamefont{G.}~\bibnamefont{Hautier}},
  \bibinfo{author}{\bibfnamefont{M.}~\bibnamefont{Kocher}},
  \bibinfo{author}{\bibfnamefont{S.}~\bibnamefont{Cholia}},
  \bibinfo{author}{\bibfnamefont{D.}~\bibnamefont{Gunter}},
  \bibinfo{author}{\bibfnamefont{V.~L.} \bibnamefont{Chevrier}},
  \bibinfo{author}{\bibfnamefont{K.~A.} \bibnamefont{Persson}},
  \bibnamefont{and} \bibinfo{author}{\bibfnamefont{G.}~\bibnamefont{Ceder}},
  \bibinfo{journal}{Computational Materials Science}
  \textbf{\bibinfo{volume}{68}}, \bibinfo{pages}{314 } (\bibinfo{year}{2013}),
  ISSN \bibinfo{issn}{0927-0256},
  \urlprefix\url{http://www.sciencedirect.com/science/article/pii/S0927025612006295}.

\bibitem[{\citenamefont{Zhang and Zhang}(2017)}]{AELAS}
\bibinfo{author}{\bibfnamefont{S.}~\bibnamefont{Zhang}} \bibnamefont{and}
  \bibinfo{author}{\bibfnamefont{R.}~\bibnamefont{Zhang}},
  \bibinfo{journal}{Computer Physics Communications}
  \textbf{\bibinfo{volume}{220}}, \bibinfo{pages}{403 } (\bibinfo{year}{2017}),
  ISSN \bibinfo{issn}{0010-4655},
  \urlprefix\url{http://www.sciencedirect.com/science/article/pii/S0010465517302400}.

\bibitem[{\citenamefont{Kumar et~al.}(2014)\citenamefont{Kumar, Kashyap,
  Balamurugan, Shield, Sellmyer, and Skomski}}]{Kumar_2014}
\bibinfo{author}{\bibfnamefont{P.}~\bibnamefont{Kumar}},
  \bibinfo{author}{\bibfnamefont{A.}~\bibnamefont{Kashyap}},
  \bibinfo{author}{\bibfnamefont{B.}~\bibnamefont{Balamurugan}},
  \bibinfo{author}{\bibfnamefont{J.~E.} \bibnamefont{Shield}},
  \bibinfo{author}{\bibfnamefont{D.~J.} \bibnamefont{Sellmyer}},
  \bibnamefont{and} \bibinfo{author}{\bibfnamefont{R.}~\bibnamefont{Skomski}},
  \bibinfo{journal}{Journal of Physics: Condensed Matter}
  \textbf{\bibinfo{volume}{26}}, \bibinfo{pages}{064209}
  (\bibinfo{year}{2014}),
  \urlprefix\url{https://doi.org/10.1088%2F0953-8984%2F26%2F6%2F064209}.

\end{thebibliography}
 

\end{document}